\documentclass[pre,twocolumn,superscriptaddress,longbibliography,showpacs,nofootinbib]{revtex4-1}

\usepackage{bbm}
\usepackage[centertags]{amsmath}
\usepackage{amsfonts}
\usepackage{amssymb}
\usepackage{amsthm}
\usepackage{amscd}
\usepackage{graphicx}
\usepackage{pgf}

\usepackage{times}
\usepackage{float}

\definecolor{linkcolor}{rgb}{0,0,0.8} %hyperlink
\usepackage[pdftex,colorlinks=true, pdfstartview=FitV, linkcolor= linkcolor, citecolor= linkcolor, urlcolor= linkcolor, hyperindex=true,hyperfigures=true]{hyperref} %hyperlink% 

\newcommand{\mean}[1]{{\left< #1 \right>}}

   \let\ep=\epsilon

\begin{document}

\title{ Life efficiency does not always increase with the dissipation rate}

\author{Marco Baiesi}
\affiliation{
Department of Physics and Astronomy, University of Padova, 
Via Marzolo 8, I-35131 Padova, Italy
}
\affiliation{
INFN, Sezione di Padova, Via Marzolo 8, I-35131 Padova, Italy
}
\author{Christian Maes}
\affiliation{Instituut voor Theoretische Fysica, KU Leuven, Belgium}
\email{christian.maes@kuleuven.be}

\begin{abstract}
There does not exist a general positive correlation between important life-supporting properties and the entropy production rate.  The simple reason is that nondissipative and time-symmetric kinetic aspects are also relevant for establishing optimal functioning. In fact those aspects are even crucial in the nonlinear regimes around equilibrium where we find biological processing on mesoscopic scales. We make these claims specific via examples of molecular motors, of circadian cycles and of sensory adaptation, whose performance in some regimes is indeed spoiled by increasing the dissipated power.  We use the relation between dissipation and the amount of time-reversal breaking to keep the discussion quantitative also in effective models where the physical entropy production is not clearly identifiable.
\end{abstract}

\maketitle

\section{Introduction}

The complex mechanisms of life cannot be sustained in thermodynamic equilibrium; they emerge only as a result of steady processes running far enough from equilibrium. Hence, it does not seem wholly unnatural to believe that life can only become better, stronger, and more robust when farther and farther from equilibrium. One standard measure of the distance from equilibrium is the dissipation rate.  We may be tempted then to expect that there exists a quite general positive correlation between dissipation rate and properties which are beneficial for life. In fact in recent decades and probably starting with the vision of life as a dissipative structure \cite{pri77nobel}, there has been a strong focus on the role of entropy production and on energy--entropy balances in the evolution and functioning of life; see {\it e.g.}~\cite{eng13,eng15,rue17,lan12,ten12,mar06}.

However, if steady dissipation as hallmark of irreversibility was the key-element for explaining the structure of life mechanisms, their stability and performance should be related systematically with the dissipation rate. For example, some types of currents are seen as oscillations, such as circadian cycles~\cite{dun99,gon02,tak16} or biorhythms.  The presence of such cycles and their period are important and must be endogenously robust. Would it help to increase the dissipation rate? (We show a novel counterexample in Sec.~\ref{bru}.)   Similarly, one may wonder whether rigidity transitions in biological tissues~\cite{bi15,bi16} are essentially steered by dissipative effects.  

In some cases we already know that increased dissipation corresponds to regimes with lower efficiency or performance.
For molecular motors~\cite{hof16,lau07,mae15}, there exist models where the efficiency of the motor was shown explicitly to decrease by driving the system farther away from equilibrium~\cite{par99}. We add a new example in Sec.~\ref{kine}.
Another case is that of kinetic proofreading~\cite{hop74}: biological error corrections serve the purpose of producing the correct population inversion with respect to the equilibrium distribution. In some models of proofreading and similar dynamics~\cite{sar13,gas16,oul17,gov14,cui17arxiv,des17arxiv}, it is clear that the selection of the ``correct or useful'' configuration statistics is not decided by the entropy production rate and may be even decreasing with it. 
Biological processes also appear to be helped sometimes by being ``jammed'' in some state~\cite{har16}, for improved stability such as in cellular physiological homeostasis. It is not clear whether such low susceptibility is better reached by increasing the dissipation rate. However, there are models of sensory adaptation in which the biological levels of specific concentrations would be destabilized by an increased dissipation~\cite{dep13, bui14} (we will touch this issue in Sec.~\ref{ada}). A larger dissipation may very well be associated with the greater possibility of establishing complex patterns far from equilibrium, but that does not seem to suffice and intermediate values of dissipation appear to be preferred in real systems.

In this paper we explain why the quality of a life-supporting process cannot depend only on the amount of dissipation. There is a good theoretical reason for not focusing entirely on entropy production when dealing with nonequilibrium systems. We know since some time that minimum and maximum entropy production principles~\cite{bru07} are in general restricted to the linear regime around equilibrium, while {\it true} stationary nonequilibrium statistics is also governed by time-symmetric kinetic aspects; see {\it e.g.} the blowtorch theorem of Landauer \cite{lan75,mae13.b}. 
A steady nonequilibrium condition for an open system is not only characterized by dissipation, but also by kinetic aspects that quantify the activity in the system and that are nondissipative by definition~\cite{lan75,lec05,mer05,gar09,gor09,bai09,bai13,lip14,bas15,mae15,mae16arxiv,mae16,jac16,hel16,fal16b,yol17}

We start in the next session by recalling the connection of entropy production with the breaking of time--reversal invariance.  This furnishes a general way to estimate the distance of a process from equilibrium. At the same time, what is complementary to entropy production can be identified with time-symmetric components and parameters in the path-probabilities and with quantities such as the dynamical activity.

We then make our case more specific by treating three examples, in Section \ref{ex}, where the distance from equilibrium is measured via suitable dissipation rates, and what is good and efficient for the life process is defined and motivated in each specific case.
We deal with models of the kinesin molecular motor~\cite{lau07}, of a circadian cycle, and of sensory adaptation~\cite{war07,lan12}, all discussed on the level of mesoscopic biophysics modeling.

In Section \ref{theo}, besides mentioning more examples, we discuss how kinetic considerations are related to nonequilibrium response and effective forces, and to how those forces are not entirely -- and sometimes entirely not -- entropic.

\section{Quantifying time (anti)symmetry}

In this section we recall how  symmetry {\it versus} antisymmetry under time--inversion leads to complementary concepts in the construction of nonequilibrium physics.  We start with dissipation as a time-antisymmetric concept,  and we end with the time-symmetric sector.

\subsection{Dissipation and distance to equilibrium} \label{ep}

Thermodynamic equilibrium for the particle density or energy profile in a macroscopic closed isolated system is obtained at the value $x^*$ whose phase space volume $W(x^*)$ (which counts the microscopic states compatible with $x^*$) is overwhelmingly larger than that of other $x$'s. One may thus quantify the departure from equilibrium via the entropy difference  $S(x^*) - S(x)$, where  $S(x) = k_B\log W(x)$.  However, a notion of distance from equilibrium based on the entropy $S(x)$ or on free energy for open systems becomes less useful when dealing with observables that depend on trajectories, such as currents or measures of dynamical activity (roughly speaking, the latter corresponds to the number of jumps between different states~\cite{lec05,mer05,gar09}).
Moreover, kinetic modeling often does not come explicitly with a thermodynamic interpretation.  These considerations, in particular, are applicable to many biological models on mesoscopic scales.

Another notion for the distance to equilibrium then may enter, which is basically telling us how large are the dissipative currents maintained in stationary nonequilibrium systems through the steady contact with different reservoirs. The corresponding mean entropy production is the total change of (equilibrium) entropy in the environment, the sum of the entropy changes in each reservoir (which is large and always in its own equilibrium), or the sum of the dissipated heat in each chemo-thermal reservoir divided by temperature~\cite{der07}.  An interesting finding of about twenty years ago is that at least under some conditions of \emph{local} detailed balance~\cite{kat84,mae03,har05,der07}, the path-dependent entropy flux as introduced above can be obtained also \emph{directly from the dynamics of the subsystem itself}; see \cite{cro98,mae99,mae00,mae03, maes03}.   Skipping the details, one result has been that the stationary entropy production per $k_B$ for a given process over time $t$ equals the relative entropy between the forward and the backward evolution probabilities,
\begin{equation}\label{rela}
\sigma\,t =  {\cal S}(P|P\theta) = \int {\cal D}[\omega]\, P[\omega]\log \frac{P[\omega]}{P[\theta\omega]}
 \end{equation}
where $\sigma$ is the mean entropy production rate.

In this formula, the formal integration goes over all possible trajectories $\omega$ of the subsystem on some level of biological or chemophysical coarse graining; $\theta \omega$ is the time-reversal of $\omega$.   As a consequence of the assumed fundamental reversibility of physical systems, when $\omega$ is an allowed trajectory, so is $\theta \omega$.  The probabilities $P[\omega]$ and $P[\theta\omega]$ are only equal in general under equilibrium.  Off-equilibrium, as for many biological processes, $P \neq P\theta$ which says that time-reversal symmetry is violated.  In the mentioned references that distinction \eqref{rela} between these two stationary path-probabilities, measuring the plausibility of a trajectory against its time-reversal was found to be coinciding with the stationary entropy production per $k_B$ when the applied modeling allows a thermodynamic identification of heat and entropy fluxes. 

On mesoscopic scales where the relevant energies are of the order of the thermal energy $k_BT$ theoretical modeling uses stochastic processes that, while case by case relevant for the discussed biophysics, do however {\it not always} provide a simple identification of the physical entropy production.  In those cases, \eqref{rela} can still be used as an estimator of the distance to equilibrium. In fact, if only as an abuse of terminology, one could very well keep calling \eqref{rela} itself the stationary entropy production per $k_B$, even in the absence of a clear thermodynamic interpretation for the model at hand.  The $\sigma$ in \eqref{rela} certainly keeps the meaning of a dissipative measure of distance away from thermal equilibrium, of course always to be understood as corresponding to a given level of coarse-graining.

\subsection{Nondissipative parameters and quantities}\label{ndp}
As a natural continuation of the previous explanations, we consider nondissipative those parameters or quantities that are time-symmetric.  
See \cite{mae17arxiv} for a recent pedagogical review.

To be specific and to introduce some of the notation that follows in the next sections, we concentrate here on mathematical modeling of an open system dynamics via a Markov jump process for which the state occupations $\rho_t(x)$ change with time following the Master equation,
\[
\dot{\rho}_t(x) = \sum_y [k(y,x)\,\rho_t(y) -k(x,y)\,\rho_t(x)]
\]
The states denoted by $x,y,\ldots$  give, for example, the position of particles or the chemomechanical configuration of a molecule, or the occupation on an energy level, etc.  The transition rates $k(x,y)\geq 0$ for the jump $x\rightarrow y$ can always be decomposed in a time-symmetric and a time-antisymmetric part,
\begin{eqnarray}\label{ps}
k(x,y) &=& \sqrt{k(x,y)k(y,x)}\,\sqrt{\frac{k(x,y)}{k(y,x)}} \nonumber\\
&\equiv& \psi(x,y)\,e^{s(x,y)/2}
\end{eqnarray}
assuming that $k(x,y)\neq 0$ iff $k(y,x) \neq 0$ to retain dynamical reversibility.  Under the same assumptions as where \eqref{rela} gives the physical dissipation, we can call
\begin{equation}\label{ldb}
s(x,y) =\log \frac{k(x,y)}{k(y,x)} = -s(y,x)
\end{equation}
the entropy change per $k_B$ in the environment over the transition $x\rightarrow y$.  Again, in many cases of physical interest, such physical interpretation follows from the dynamical reversibility of standard Hamiltonian mechanics, as referred to already above \eqref{rela}. On the other hand,
\begin{equation}\label{act}
\psi(x,y) = \sqrt{k(x,y)k(y,x)} = \psi(y,x) 
\end{equation}
is symmetric between forward and backward jumps and gives the ``width'' or ``accessibility'' of the channel. 
We call $\psi(x,y)\geq 0$ the activity parameters; they are frequencies and may depend on intensive parameters of the reservoir(s)  but also on external forces or differences in reservoir temperatures and chemical potentials, and on (free) energy barriers separating $x$ from $y$.  

A {\em nondissipative effect} occurs when the relative strength or nature of the $\psi(x,y)$ changes the nonequilibrium condition, in particular through their variation with the external field. Of course, the dissipation $\sigma$ in \eqref{rela} also depends on these activity parameters, but it is the fact that there is no potential $\cal G$ for which $s(x,y) = {\cal G}(x)- {\cal G}(y)$ for all $(x,y)$, which makes $\sigma\neq 0$.

A second class of nondissipative effects arise from the role played by time-symmetric path-observables.  In the notation of \eqref{rela} we would be speaking about observables $O(\omega)$ which are function of the trajectory over time-interval $[0,t]$ and are invariant under time-reversal $\theta$, i.e.,~$O(\omega) = O(\theta\omega)$ .  Examples are even powers of particle or energy currents, or the number of jumps in that time-interval (which is a measure of dynamical activity~\cite{lec05,mer05,gar09}), or the residence time in a certain state or collection of states; the value of each of those path-dependent quantities does not change when playing the movie of the trajectory backward.

\section{(Counter)examples}\label{ex}

There is no simple or universal definition of quality of a biological process, while, following the previous section, entropy production and dissipation can be well defined.  We thus need specific processes and models, and point to relevant nondissipative features for the intuitive well-being of the biological performance.
We are then ready for looking at three quite different models, with the aim of testing in these specific instances the metabiological hypothesis that dissipation is pushing the performances of life processes and hence that the more one dissipates, the better it is. The models provide counterexamples to that idea.  Each time we find parameters under which the entropy production and the performance are moving in opposite direction.

\subsection{Efficiency of molecular motors}\label{kine}

Upper bounds on motor efficiency in general follow from lower bounds on entropy production rate; see {\it e.g.} \cite{bok09}.  Here we consider the model of kinesin motion described in Ref.~\cite{lau07} (where one can find all the details) and we use it to
show that the most efficient pulling of a molecular cargo takes place when the availability of ATP -- the fuel of our motor -- is at intermediate physiological values, where dissipation is not maximal.

\begin{figure}[!tb]
\begin{center}
\includegraphics[angle=0,width=6cm]{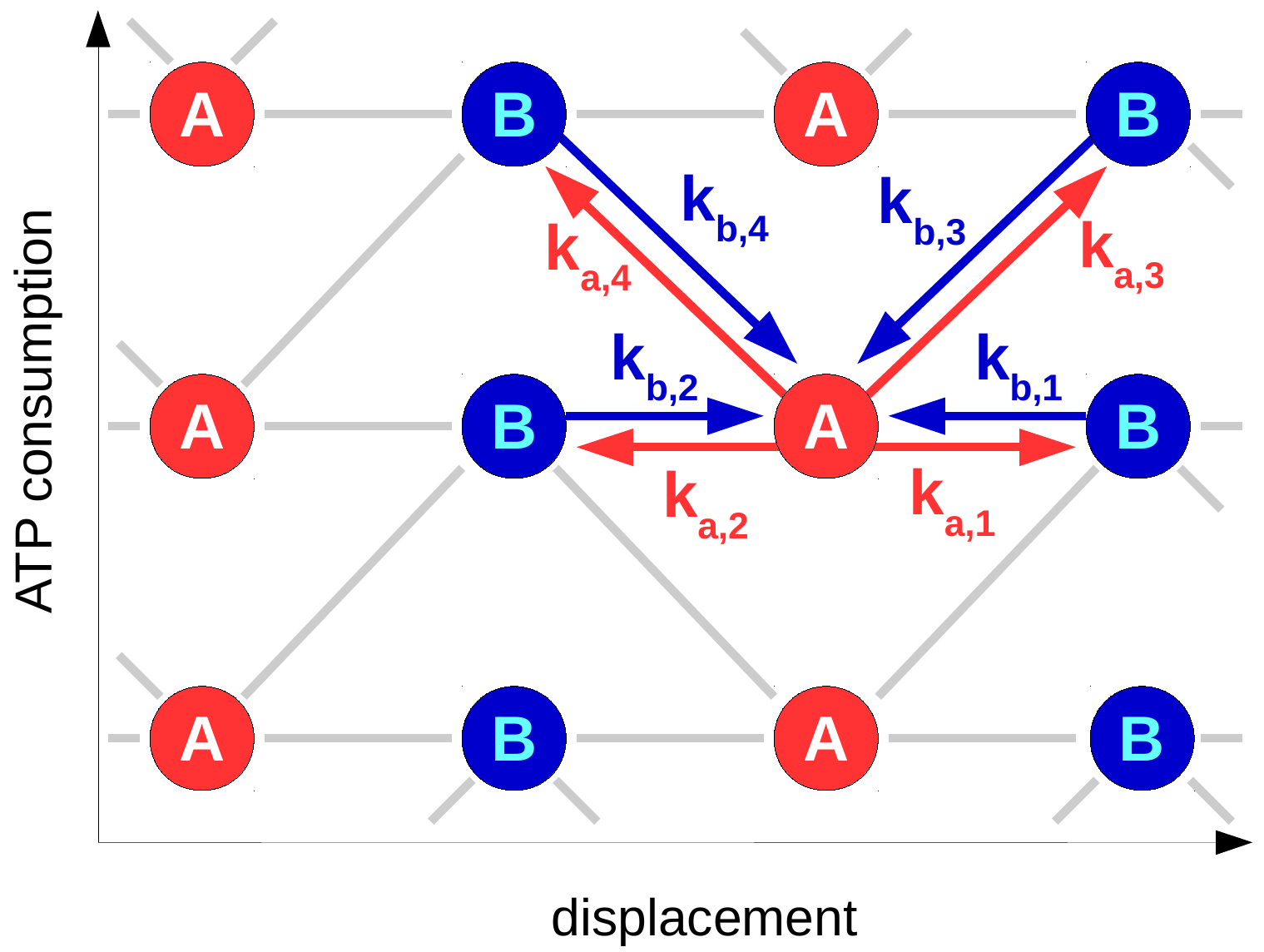}
\end{center}
\caption{
Portion of the infinite network of states for the kinesin model.
Lines indicate allowed transitions between the two types of state  (``A'' and ``B'') displayed as a function of the accumulated displacement of the molecular motor and of the number of consumed ATPs. We have also specified the transition rates from ($k_a$'s) and to ($k_b$'s) a particular state ``A'', see the text for more details.}
\label{fig:kin}
\end{figure}

The motor can be either in a state ``A'' or in an activated state ``B''. The transition between ``A'' and ``B'' can take place through thermal fluctuations (horizontal transitions in the scheme of Fig.~\ref{fig:kin}) or by ATP consumption/release (diagonal transitions). In Fig.~\ref{fig:kin} each state is displayed {\it vs.}~the position along the microtubule over which the motor is stepping and {\it vs.}~the amount of consumed ATP. A motor full step $2 d \approx 8$~nm, corresponding to the horizontal gap between two ``A'' states in Fig.~\ref{fig:kin}, usually displaces the kinesin on the right, even if there is a load that imposes an external force $F_e<0$, i.e., directed on the left. 

%%%%%%%%%%%%%%%%%%%%%%%%%%%%%%%%%%%%%%%%%%%%%%%%%%%%%%%%%%%%%%%%%%%
%%%%%%%%%%%%%%%%%%%%%%%%%%%%%%%%%%%%%%%%%%%%%%%%%%%%%%%%%%%%%%%%%%%

By plugging in the values of parameters from the fit to experimental data in~\cite{lau07}, for $f=F_e d/(k_B T) = -4$ (in modulus below the value $f\approx -4.87$ obtained with the stalling force $F_e\approx -5$ pN of kinesin~\cite{car06}), we get the rates (in $s^{-1}$)
\begin{align}
k_{a,1} &  \simeq 2.60\times 10^{-5} &
k_{b,1} &  \simeq 1.85 \nonumber\\ 
k_{a,2} &  \simeq 3.30 & 
k_{b,2} &  \simeq 78.5 \nonumber\\ 
k_{a,3} &  \simeq 0.593 \times \textrm{[ATP]} & 
k_{b,3} &  \simeq 0.301 \nonumber\\ 
k_{a,4} &  \simeq 0.00556 \times \textrm{[ATP]} & 
k_{b,4} &  \simeq 9.44\times10^{-7} \nonumber
\end{align}
where concentration [ATP] is in mM units.
Clearly transitions with rates $k_{a,4}$, $k_{b,4}$, and $k_{a,1}$ are suppressed; the motor usually repeats multiple jumps along the transition $2$ till the transition with $k_{a,3}$ is followed.

%%%%%%%%%%%%%%%%%%%%%%%%%%%%%%%%%%%%%%%%%%%%%%%%%%%%%%%%%%%%%%%%%%%
\begin{figure}[!tb]
\begin{center}
\includegraphics[angle=0,width=7.5cm]{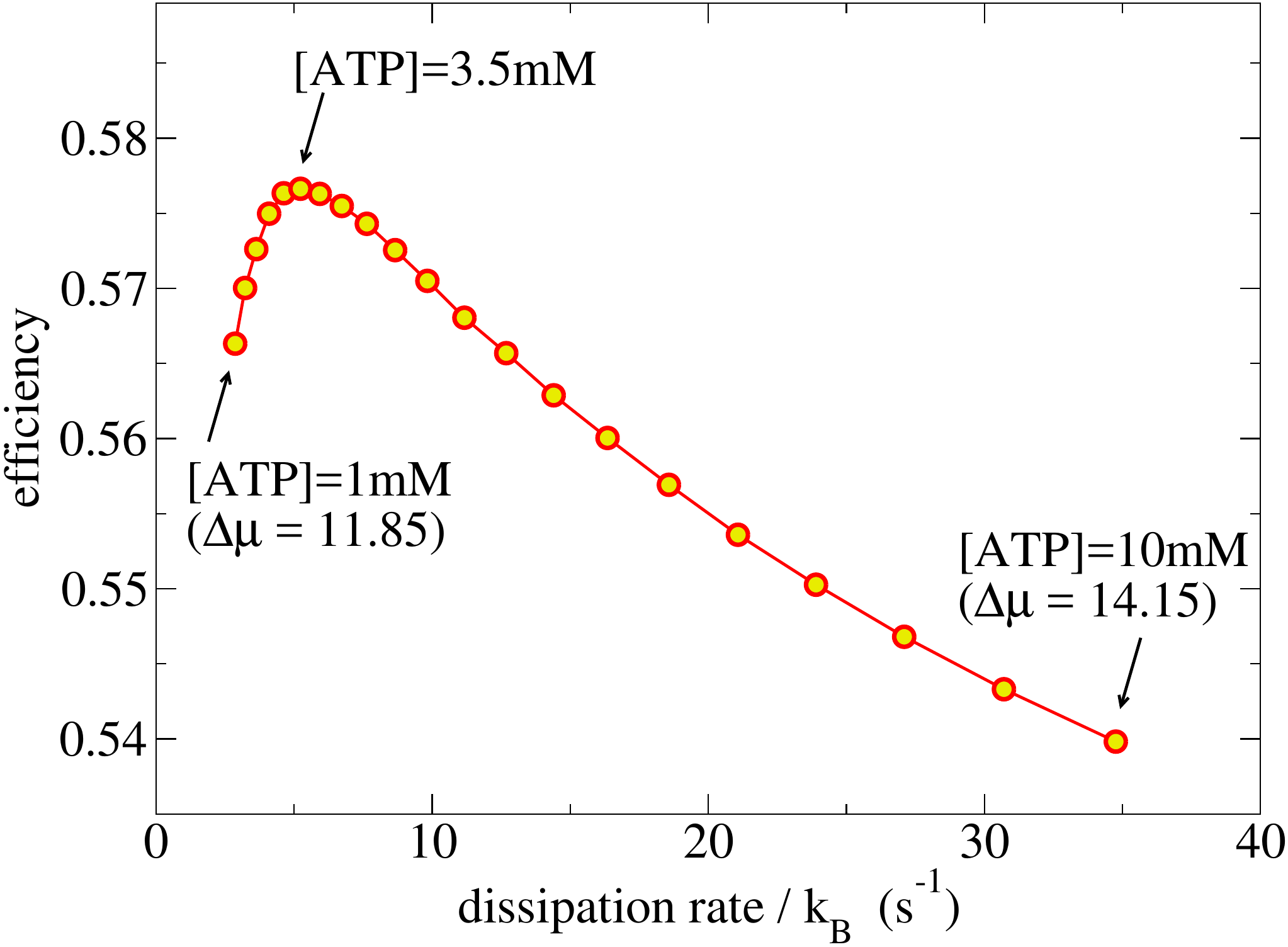}
\end{center}
\caption{Kinesin efficiency {\it vs.}~dissipation rate. Points are for values of $\Delta\mu=\log(k_0 \textrm{[ATP]})$ equally spaced between $\Delta\mu=11.85$ (corresponding to [ATP]=$1$ mM) and $\Delta\mu=14.15$ ([ATP]=$10$ mM). The maximum efficiency is obtained for [ATP]$\approx 3.5$ mM.
\label{fig:kin2}}
\end{figure}
%%%%%%%%%%%%%%%%%%%%%%%%%%%%%%%%%%%%%%%%%%%%%%%%%%%%%%%%%%%%%%%%%%%

%%%%%%%%%%%%%%%%%%%%%%%%%%%%%%%%%%%%%%%%%%%%%%%%%%%%%%%%%%%%%%%%%%%
\begin{figure*}[!t]
\begin{center}
\includegraphics[angle=0,width=4.5cm]{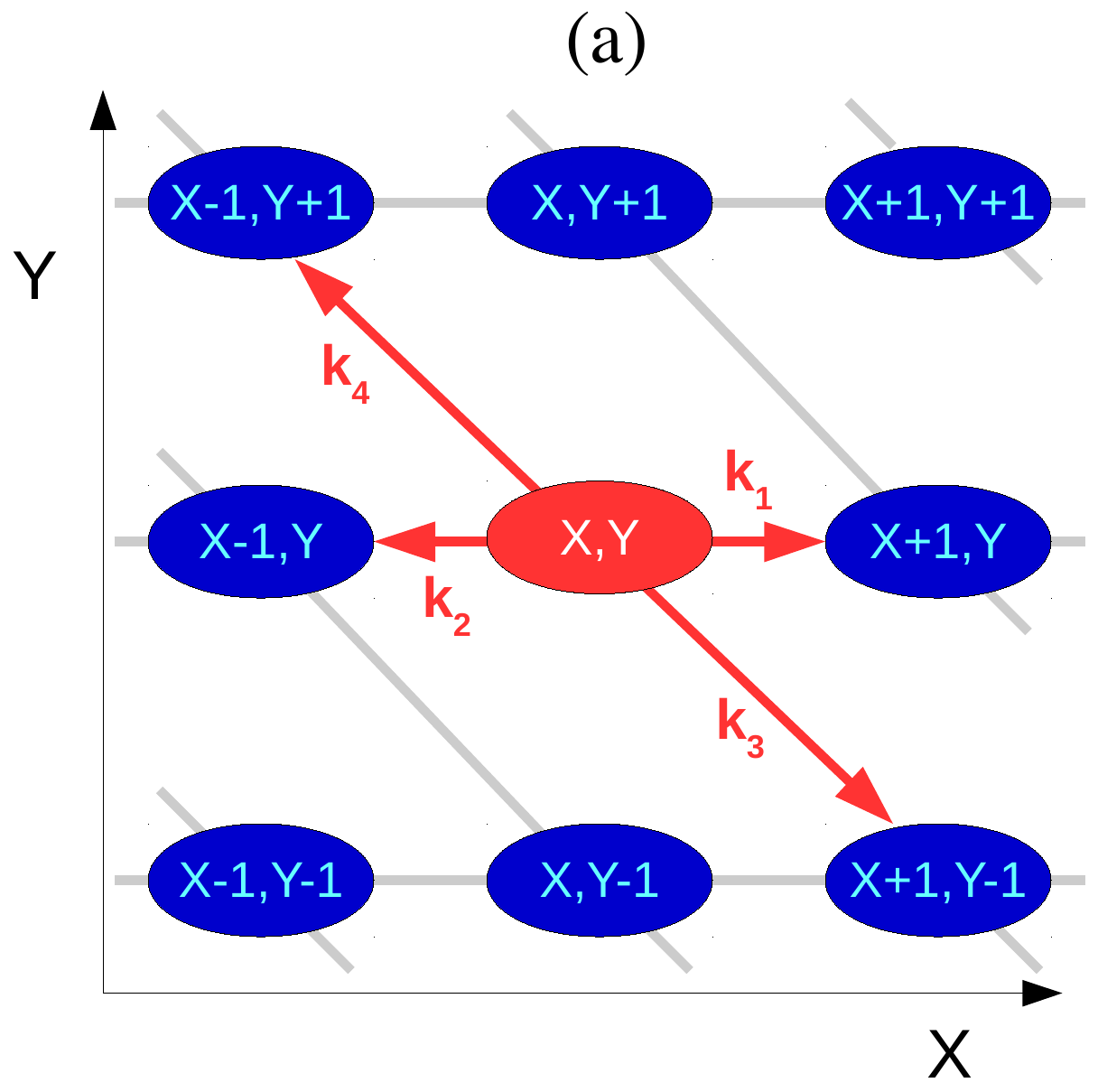}
\includegraphics[angle=0,width=11cm]{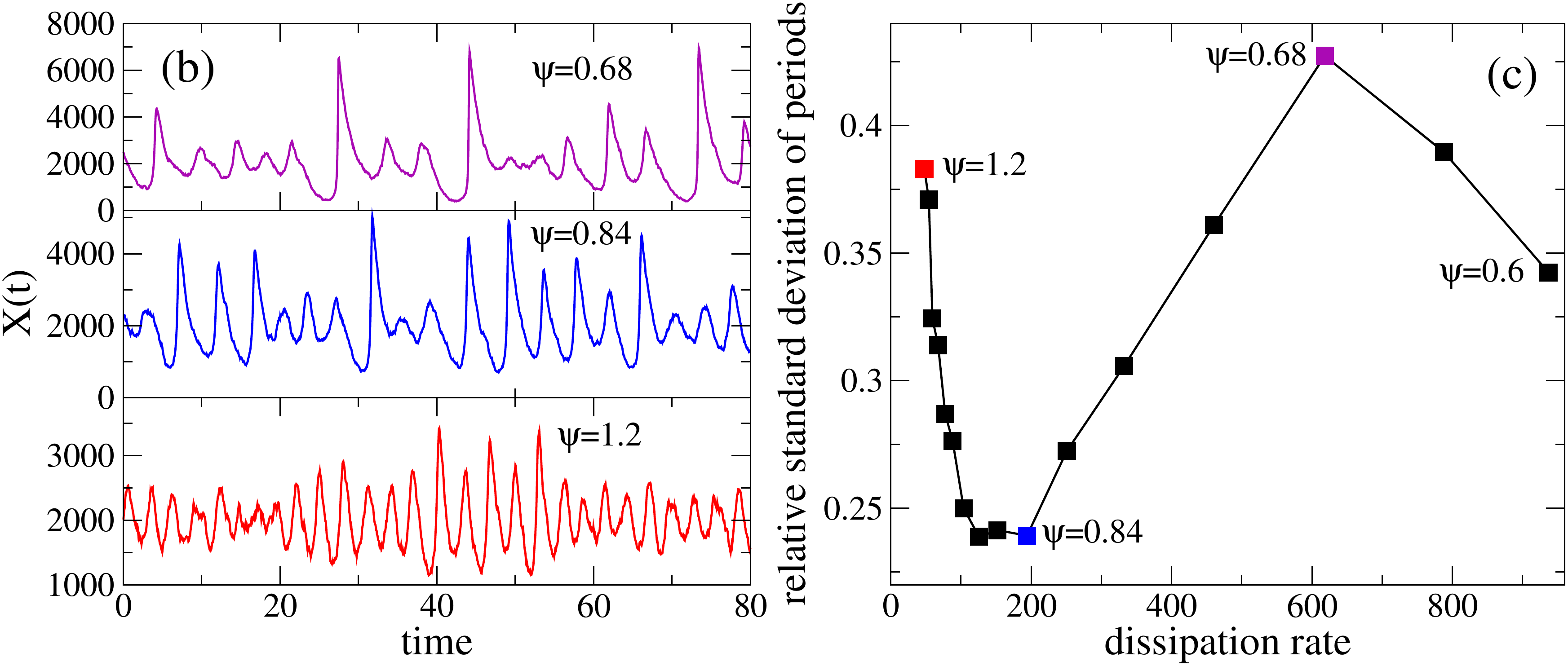}
\end{center}
\caption{
(a) Sketch of the allowed transitions from a state $x=(X,Y)$ of the Brusselator and their rates.
(b) Time series of the $X$ variable for three different values of the activity parameter $\psi$. 
(c) The relative standard deviation of periods {\it vs.}~entropy production rate, for the Brusselator. Squares from right to left are parametrized by values of the activity parameter $\psi=0.6$, $0.64,\ldots$, $1.2$ . In this range the best regularity in periods is achieved around the value $\psi=0.84$.
\label{fig:bru}}
\end{figure*} 
%%%%%%%%%%%%%%%%%%%%%%%%%%%%%%%%%%%%%%%%%%%%%%%%%%%%%%%%%%%%%%%%%%%

We have varied the ATP concentration in the typical physiological range, from $1$ mM up to $10$ mM, to check if the performance of the motor gets better with increasing dissipation. In the normalized units used above and in Fig.~\ref{fig:kin}, where states should be thought on a square lattice with edges of unit side, the motor average velocity $v$ equals the horizontal displacement per unit time and the average ATP dissipation rate $r$ is the mean vertical displacement per unit time.
The quality of the motor's performance is quantified by its efficiency
\begin{equation}
\eta \equiv - \frac{ f v} {r \Delta\mu}
\end{equation}
which is the ratio of dissipated power $-f v$ and input power $r \Delta\mu$ (with $\Delta\mu = \log(k_0 [ATP])$ and $k_0=1.4\times 10^{-5}$ mM$^{-1}$, see~\cite{lau07}).
Dissipation is quantified by the mean entropy production per unit time, obtained by averaging in time the  entropic  contribution from jumps, such as $s_3(x=A,y=B) = \log k_{a,3}/k_{b,3}$.

In Fig.~\ref{fig:kin2} we see that the efficiency is not a monotone function of the dissipation but rather finds a maximum at intermediate physiological conditions of ATP concentration, a finding likely pointing to a natural selection mechanism that led kinesin to operate in optimal conditions. For our point, we note that the performance gets worse if one increases too much the ATP concentration and consequently the dissipation of the system.

\subsection{Regularity of circadian clocks}\label{bru}

To better couple with the environment, for an organism it is often convenient to have a physiological state with variables ({\it e.g.}, enzyme concentrations) that follow an oscillation of $24$ hours~\cite{dun99,gon02,tak16}. A circadian clock is present if there is an  endogenous component in this oscillation, namely the cycle remains rather stable even in the absence of external daily stimuli. 
To simulate a circadian cycle, we consider the so called Brusselator~\cite{pri68,tys08}, which was invented to model well known chemical periodicity, such as in the Belousov--Zhabotinsky reaction.
In order to emphasize the role of the time-symmetric components $\psi(x,y)$ in front of the jumping rates \eqref{ps}, we add a parameter $\psi$ before a pair of forward-backward transition rates to modulate the volume of jumps along that direction and show that the entropy production decreases with $\psi$, while the quality of the clock becomes better at intermediate values of $\psi$, as detailed in the following.

A state $x=(X,Y)$ of the system is represented by the numbers $X$ and $Y$ of molecules of two different species, hence $x$ belongs to the positive quadrant of the square lattice. The stochastic version of the Brusselator, translating the original deterministic dynamics~\cite{pri68} into a Markov jump process, takes into account the finite size of the system via a ``volume'' $\Omega$ \cite{gon02} that appears in the rates of allowed transitions. Following the sketch of Fig.~\ref{fig:bru}(a), these rates are 
\begin{align}
k_1(x) &= \psi a \Omega  &\textrm{for}\quad (X,Y)\to (X+1,Y) \nonumber\\
k_2(x) &= \psi X         &\textrm{for}\quad (X,Y)\to (X-1,Y) \nonumber\\
k_3(x) &= \frac 1{\Omega^2}\,X(X-1) Y   &\textrm{for}\quad (X,Y)\to (X+1,Y-1) \nonumber\\
k_4(x) &= b X  &\textrm{for}\quad (X,Y)\to (X-1,Y+1) \nonumber\\
\end{align}
We will set  $\Omega=1000$, $a=2$, $b=5$. This value of $b$ is large enough to see the appearance of oscillations with a limit cycle around which the stochastic dynamics settles quite quickly.

We can use the relative entropy between forward and backward trajectories for estimating the mean entropy production of the Brusselator, as explained in Section \ref{ep}. The relevant entropy fluxes are
\begin{align}
s(X,Y \rightarrow  X+1,Y) &= \log\frac{\Omega a}{X+1} \label{bru_s1} ,\\
s(X,Y \rightarrow  X-1,Y+1) &= \log\frac{{\Omega^2} b X}{(X-1)(X-2)(Y+1)} \label{bru_s2} .
\end{align}
 There is  a small technical problem with the transition $(2,Y  \rightarrow  1,Y+1)$ (rate $\psi b\ne 0$), whose inverse $(1,Y+1  \rightarrow 2,Y)$ is forbidden. We can however locally modify the scheme and associate an arbitrary small rate to that reaction without changing the main analysis. Thus, (\ref{bru_s2}) is changed into $1/\ep$ for $X=1,2$.
We note again that the activity parameter $\psi$ of the jump rates has disappeared from these entropy productions. Fig.~\ref{fig:bru}(b) shows examples of time series of $X$ (those of $Y$ are similar) obtained for three different values of $\psi$.

The quality of the Brusselator clock is estimated from the distribution of the periods, specifically from its standard deviation normalized by the mean period, i.e.~the periods relative standard deviation.
To identify full cycles and hence their periods, first we smoothen the time series of the $X$ variable by averaging $X$ in time steps $\Delta t = 0.05$. Then, we estimate the interoccurrence times between subsequent main peaks above the threshold $X_0 = 2000$. This threshold should be re-crossed from below at least after a time $5\Delta t$ before restarting with a new peak identification.  We have tested that  values of $X_0 \in[1900,2100]$ give similar results. Moreover, by visual inspection we checked that the peak recognition works well, especially for $\psi<1$.

The values $\psi=0.68, 0.84$, and $1.2$ used in Fig.~\ref{fig:bru}(b) characterize a non-monotonic trend of the periods' relative standard deviation, as shown in Fig.~\ref{fig:bru}(c), where we see that the best performance is obtained around $\psi=0.84$, an intermediate value if the range $\psi \in [0.6, 1.2]$ is considered. By comparing specifically the two series for  $\psi=0.84$ (lower dissipation rate) and $\psi=0.68$ (higher dissipation rate), shown in Fig.~\ref{fig:bru}(b), we see that the decays after peaks in the $\psi=0.84$ time series have a more regular pace. Thus, there is a region where the increase of entropy production rate would lead to a higher volatility of periods.
To summarize, we again find no positive correlation between the quality of the system and its dissipation rate.

\subsection{Precision of sensory adaptation}\label{ada}

We consider a minimum feedback network underlying many sensory adaptation systems~\cite{lan12}.
A level of time-dependent ``output activity'' $a(t)$ (not to be confused with the dynamical activity in stochastic processes) is maintained around a physiological level $a_0$ by means of a feedback mechanism: a buffer variable $m(t)$ reacts to variations of an external stimulus $s$ and, eventually, its feedback maintains the level of $a$ close to the optimal $a_0$. 
We are interested to see if on average $a$ remains closer to $a_0$ when dissipation is higher. Again, in the following we show that better performance in general is not associated with higher dissipation.

The whole system represents a small fluctuating ensemble of molecules, which is conveniently described at a mesoscopic level by diffusion equations (see~\cite{wan15} for the jump process version of the model). In the notation of Ref.~\cite{lan12}, these are
\begin{equation}\label{sa}
\begin{split}
  \dot{a} &= F_a + \sqrt{2 \Delta_a}\,\xi^a(t) \\
  \dot{m} &= F_m + \sqrt{2 \Delta_m}\,\xi^m(t)
\end{split}
\end{equation}
 with ``forces''
\begin{equation}
\begin{split}
F_a &= -\omega_a  [a - G(s,m)] \\
F_m &= -\omega_m (a - a_0) \left[ \beta - (1 - \beta) \,C\, \partial_m G(s,m)  \right]
\end{split}
\end{equation}
that represent biochemical interactions at a coarse-grained level.
For the $G(s,m)$ function we take the Michaelis-Menten form $G(s,m) =(1+ s e^{-2m})^{-1}$,
with $\partial_s G<0$ and $\partial_m G>0$ as required  for a negative feedback mechanism.
The dynamics is stochastic via the white noise terms $\xi_a,\xi_m$ with amplitudes $\Delta_a,\Delta_m$, respectively. 
Following~\cite{lan12}, $C = \frac{\Delta_m / \omega_m}{\Delta_a / \omega_a}$ so that for $\beta=0$ there is detailed balance for potential $V(a,s,m) = \frac{\omega_a}{\Delta_a}\left[ \frac{1}{2}a^2 - (a-a_0)G(s,m)\right]$.
Thus, $\beta$ parametrizes the nonequilibrium component in the force $F_m$ and the system is characterized by a nontrivial feedback dynamics for large enough $\beta$, which leads $a$ to float around $a_0$.
Indeed, in the deterministic version (deleting the white noise in \eqref{sa}) there is a fixed point, $(a_0,m^*_s)$  with $G(s,m^*_s)=a_0$, which is stable when
\begin{equation}
  \beta > \beta_c = \left. \frac{C\partial_m G(s,m)}{1+C\partial_m G(s,m)} \right|_{{m = m^*}} 
\end{equation}
Note that $a_0$ in the stable fixed point does not depend on the stimulus $s$.

%%%%%%%%%%%%%%%%%%%%%%%%%%%%%%%%%%%%%%%%%%%%%%%%%%%%%%%%%%%%%%%%%%%
\begin{figure}[!t]
\begin{center}
\includegraphics[angle=0,width=8.2cm]{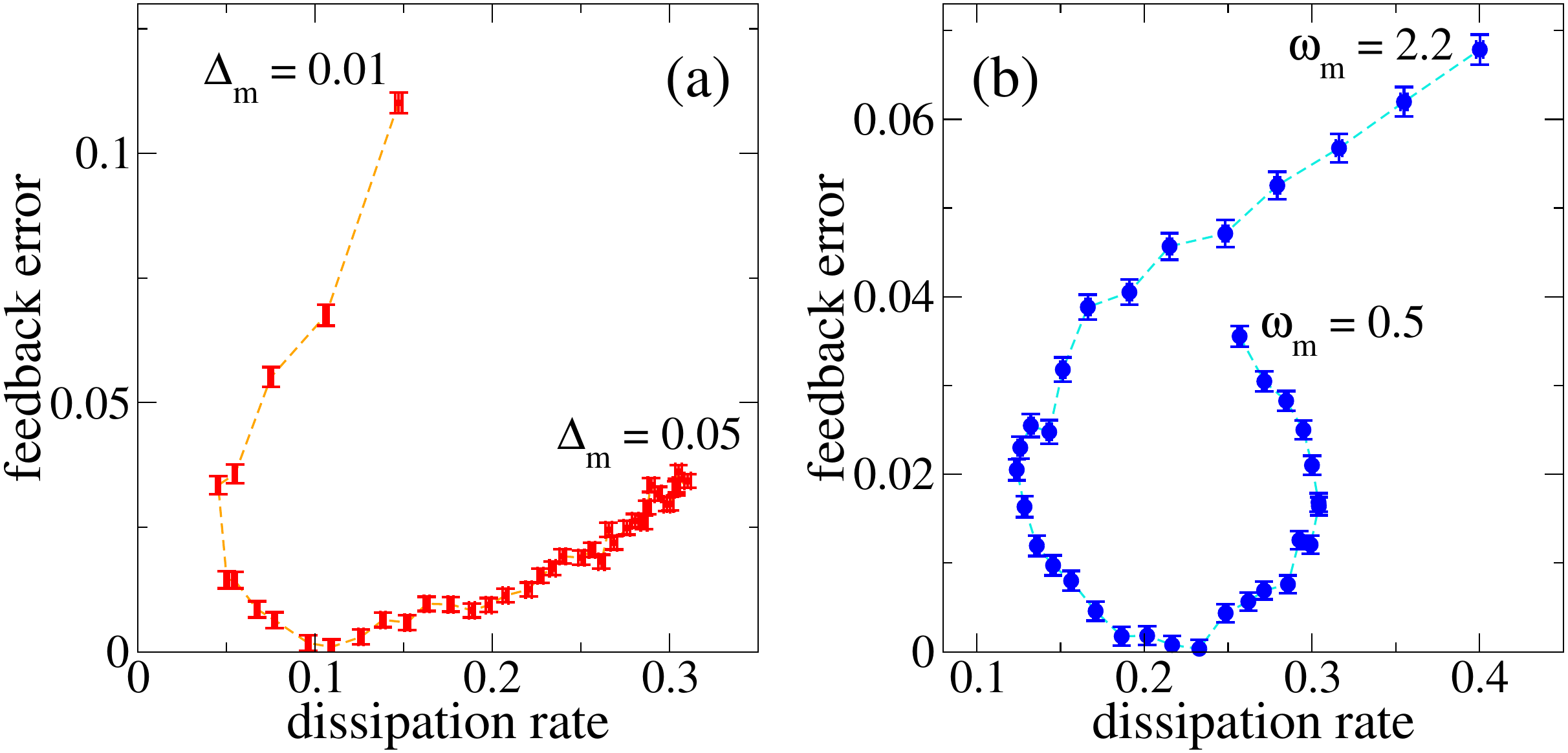}
\end{center}
\caption{
(a) Deviation of the average activity $\mean{a}$ from ideal $a_0 = 0.6$ ({\it i.e.}, ``error'' $|\mean{a}-a_0|$) plotted {\it vs.}\ the dissipation rate $\sigma$ for different values of the noise amplitude of the control variable, $\Delta_m=0.01,0.011,\ldots, 0.05$. We see that the quality of the feedback is optimal for an intermediate value of $\Delta_m$ and of the dissipation rate, while low or high $\Delta_m$ produce a higher error, which does not correlate positively with the dissipation rate.
Other parameters are $\beta=0.6$, $\omega_a=10$, $\omega_m=0.6$ (note that the feedback variable $m$ has $\omega_m \ll \omega_a$), $\Delta_a=1$, $s=5$. 
With these parameters we have $\beta_c \simeq 0.07$ and hence $\beta>\beta_c$ brings the system to the nontrivial fixed point out of equilibrium.
(b) Almost the same parameters, with fixed $\Delta_m=0.04$ and different $\omega_m=0.5,0.055,\ldots, 2.2$. Also in this example there is no general correlation between the dissipation and the quality of the feedback.
\label{fig:adapt}}
\end{figure}
%%%%%%%%%%%%%%%%%%%%%%%%%%%%%%%%%%%%%%%%%%%%%%%%%%%%%%%%%%%%%%%%%%%

The noisy dynamics (\ref{sa}) brings the system to fluctuate slightly off the fixed point and leads to a mean dissipation rate $\sigma \equiv \mean{F_a \circ d a /\Delta_a + F_m \circ d m / \Delta_m} / d t$ (products in the Stratonovich sense~\cite{sek10} and the statistical average in the steady process is denoted  by $\mean{\cdot}$).

To measure the quality of the adaptation~\cite{lan12}, one looks at the deviation $|\mean{a}-a_0|$ for cases where $\beta > \beta_c$. The $|\mean{a}-a_0|$ is a sort of error of the feedback mechanism, hence smaller $|\mean{a}-a_0|$ values indicate good adaptability. The question is whether the feedback gets better by decreasing noise amplitudes or by increasing the feedback rate $\omega_m$, and we are interested to know how better adaptation is correlated with the dissipation rate.  That was also the major question in~\cite{lan12}, where it was concluded that their study ``reveals a general relation among energy dissipation rate, adaptation speed and the maximum adaptation accuracy.''  The mathematical proportionality (equation (5) in~\cite{lan12}) between dissipation rate and adaptability is however not convincing without a more general study of the factor of proportionality.

In Figure~\ref{fig:adapt}(a) we show an example in which we vary the amplitude of the noise on the feedback variable $m$  and we clearly see that the error $|\mean{a}-a_0|$ has no general correlation with the dissipation rate. The same is true for variations of $\omega_m$, see Figure~\ref{fig:adapt}(b). These examples show that optimal points for the feedback mechanism do not correspond to maxima of the entropy production rate.

\section{Discussion on nondissipative effects}\label{theo}

A possible reason why dissipation or entropy production continue so often to play a central role in foundational discussions on life-functioning and nonequilibrium physics is  the wide appreciation and familiarity with irreversible thermodynamics, where local equilibrium and linear force-current relations constitute the usual assumptions.  Moreover, in or near equilibrium, the one and same entropy uniquely relates to heat capacity, density of states, the H-theorem, the fluctuation-dissipation theorem, thermodynamic forces, and more.    In recent years, however, it has often been emphasized that in true nonequilibrium regimes the Boltzmann-Clausius correspondence between heat and degeneracy, or between thermodynamic potential and fluctuations gets broken. From second order onward in any driving, the nonequilibrium statistics is described dynamically both by a dissipative, time-antisymmetric quantity (entropy production) \`and by kinetic time-symmetric estimators, sometimes called dynamical activity~\cite{lec05,mer05,gar09}, traffic~\cite{mae08a}, or frenesy~\cite{bai13,bas15} (see also \cite{mae16arxiv}).

 The role of time-symmetric activity becomes evident in a version of linear response~\cite{bai09,bai13,bas15,fal16b,yol17} in which the differential mobility, the change of a current $J$ over $[0,t]$ due to a  variation of a parameter or to a perturbation,
\begin{equation}\label{jres}
\partial \langle J\rangle =  \frac 1 2 \mean{{\cal S}(\omega) J(\omega)} - \mean{K(\omega) J(\omega)}
\end{equation}
is written as a difference between two terms, both being dynamical correlations in the unperturbed steady nonequilibrium with $\langle\cdot\rangle$ averaging over the possible trajectories $\omega$. Here the ${\cal S}(\omega)$ is the path-dependent entropy flux due to the perturbation, and $K(\omega)$ is the path-dependent time-symmetric dynamical activity ({\it e.g.}, including the changes in residence times or in undirected currents caused by the perturbation).  Very often the relevant current is itself proportional to ${\cal S}$.  Then, for $J\propto {\cal S}$ in \eqref{jres}, the first correlation $\mean{{\cal S}^2} > 0$ is certainly positive, and it is only possible to cancel it by the second correlation $\mean{K\, {\cal S}}$ when far enough from equilibrium. Such a cancellation is indeed impossible in equilibrium, where always $\mean{K(\omega) J(\omega)}_\text{eq} = 0 = \mean{K(\omega) {\cal S}(\omega)}_\text{eq} $ by time-reversal symmetry~\cite{bai13,bas15}.  That is a typical example of how, through the presence of nonzero dissipation in nonequilibrium, the time-symmetric sector (in terms of $K$) becomes relevant and creates important possibilities in bio-processing.  E.g., to reach a homeostatic regime,  biological processes might exploit a stalling of relevant quantities/currents to external stimuli.
The physics of glasses, in which caging is the important effect, and the corresponding studies of changes in dynamical activity have also become biologically relevant \cite{bi16}.
In a different context, when dealing with driven particles, the stalling might be the point at the onset of a regime of negative differential mobility~\cite{bas15,bae13,bai15,sar16}. 

Physics-oriented studies of adaptability may also consider response relations like \eqref{jres}, in which the observable $\cal O$ is now not a current but rather a state function, like in Section \ref{ada}, and where again the second term in \eqref{jres} now in the form $\langle {\cal O}\,K\rangle$ makes the essential difference from the usual fluctuation--dissipation relation (a fluctuation-dissipation relation reproducing the standard equilibrium version can be found for stalled currents in \cite{alt16}) enabling for example to decrease the susceptibility.  Similar relations and considerations apply starting from second order around equilibrium \cite{bas15.c}.

In a recent work~\cite{gas16} we may find other results supporting our point, which were however presented without the emphasis of the present paper. Some models of exonuclease proofreading and  biological error-correction were investigated~\cite{gas16} and the error probability was seen to increase together with the entropy production rate as a function of growing nucleotide concentration in physiological regimes.  It is the dependence of the activity parameters on driving that allows the error probability to decrease with the nucleotide concentration.

Myosin V, a molecular motor, is another example where the role of time-symmetric quantities emerges  clearly~\cite{mae15}. Tuning the activity parameters in the corresponding jump rates, it was shown that the motion of myosin V can even change direction if the volume of transitions between specific states is changed \cite{mae15,mae16arxiv, mae16}. This has nothing to do with entropy production, as the bi-directional increase of jumps between say states $x$ and $y$ (the {\it traffic}) is governed by the activity parameter $\psi(x,y) = \psi(y,x)$ defined in Sec.~\ref{ep}. Indeed, the transition frequency along a given channel is the main factor determining the direction of the molecular motor.

The above situations are much different from the macroscopic effect of currents (and power) increasing with a driving potential (as in Ohm's law), to which we are acquainted with near-equilibrium linear response. Far from equilibrium, induced forces are no longer minus the gradient of a thermodynamic potential, and they can realize motion or increased stability of fixed points only by the combination of entropic and frenetic effects~\cite{bas16}.
Near equilibrium the entropy production is just quadratic in the current, but that may change drastically farther from equilibrium. That appears again in recent studies of thermodynamic uncertainty relations~\cite{bar15,pol16,pie17,pig17arxiv,hor17,sei17arxiv}, concentrating on lower bounds for the entropy production rate and giving interesting refinements to the positivity of entropy production or to Carnot efficiency, see~\cite{hor17arxiv} and references therein, in particular \cite{mae17arxiv} for an interpretation of lower bounds on the dissipation rate.  One should not forget that quadratic lower bounds for the dissipation rate, in terms of currents, are not at all sharp in the nonlinear force-current regimes. For instance, the efficiency of the kinesin model discussed in Sec.~\ref{kine} is well below the upper limit given by the thermodynamic uncertainty relation~\cite{sei17arxiv}.

 As a simple mathematical illustration of the possible discrepancy between high dissipation and low current, we consider a one-dimensional walker on $x\in{\mathbb Z}$ with rates to jump to the right $k(x,x+1 ) = \psi(E)e^{E/2}$ and to the left $k(x,x-1) = \psi(E)\, e^{-E/2}$.  We use $E$ to denote the Joule heating caused by dissipating the external work done on the particle, reduced to dimensionless units.  For short, $E$ refers to a driving field and the escape rate
 \[
 k(x,x+1) + k(x,x-1) = 2\psi(E)\cosh (E/2)
 \] is non-monotone in $E$ when the activity parameter $\psi(E) \propto e^{-E/2}/E$ is chosen to be decreasing in $E$, a situation often occurring in the presence of obstacles.
The time-integrated current per unit time, {\it i.e.}, the number of jumps to the right minus the number of jumps to the left per unit time, is on average
\[
J(E) = 2\,\psi(E) \sinh (E/2)
\]
and the mean entropy production rate is $\sigma = E\,J(E)$.  With $\psi(E) \propto e^{-E/2}/E$ we see that $\sigma(E)$ is monotone increasing and saturating asymptotically in $E$, while $J(E)$ goes to zero as $E\uparrow \infty$ with an intermediate maximum.  It is then certainly not so that highest current is reached at highest entropy production.
Moreover, the variance of the net number of forward jumps (time-integrated current)  is completely decided by the escape rate,
\[
\langle J;J\rangle = 2t\,\psi(E)\cosh(E/2) 
\]
so that the stationary dispersion, a negative quality feature of the current,
\[
\frac{\langle J;J\rangle}{t\langle J\rangle^2} = (2\,\psi(E)\tanh (E /2)\sinh (E/2))^{-1}\stackrel{E\uparrow\infty}{\propto} E
\]
diverges (for $\psi(E) \propto e^{-E/2}/E$) where the entropy production rate reaches its maximum. Therefore, the optimal driving value for the walker is not where the mean entropy production is maximal if one wants to have a large value of the current with limited dispersion.

\section{Conclusions}

The absence of universal positive correlations between life-supporting properties and the amount of irreversibility (steady entropy production) is not truly surprising.  Trivially, in any given model of a biological system dissipative processes can be added that lower its quality.
The point of the present paper is however to give a relevant systematic and quantitative analysis, including the role of non-thermodynamic aspects.
This paper has used three models to show that more specifically: (a) kinesin in typical physiological conditions has maximum efficiency at intermediate values of ATP concentration, where the dissipation of the molecular motor is not maximum; (b) the regularity of the periods in a model of circadian clocks, the Brusselator, may become better for intermediate values of the dissipation rate; (c) a model of sensory adaptation shows no clear pattern of feedback precision improving with the entropy production rate. It does not appear generally true that ``more accurate and/or faster adaptation inevitably requires more energy dissipation per unit of time''~\cite{ten12}.

While similar claims have been made before for
several biological processes, we have presented a tool for general analysis and pointed explicitly to the role of nondissipative (time-synmmetric features).
Both dissipative and time-symmetric kinetic considerations are necessary to reach a complete picture of regimes far from equilibrium, of which biological processes are an important example.

\begin{acknowledgments}
 We are grateful to Enrico Carlon, Pierre de Buyl, Gianmaria Falasco, Pierre Gaspard, Arthur Heymans, Thomas E. Ouldridge, Alessandro Sarracino, and Shou-Wen Wang for useful discussions. M.B.~thanks the Institute for Theoretical Physics at the KU~Leuven for the hospitality.
\end{acknowledgments}

%%\bibliography{bib_noneq}

\begin{thebibliography}{73}%
\makeatletter
\providecommand \@ifxundefined [1]{%
 \@ifx{#1\undefined}
}%
\providecommand \@ifnum [1]{%
 \ifnum #1\expandafter \@firstoftwo
 \else \expandafter \@secondoftwo
 \fi
}%
\providecommand \@ifx [1]{%
 \ifx #1\expandafter \@firstoftwo
 \else \expandafter \@secondoftwo
 \fi
}%
\providecommand \natexlab [1]{#1}%
\providecommand \enquote  [1]{``#1''}%
\providecommand \bibnamefont  [1]{#1}%
\providecommand \bibfnamefont [1]{#1}%
\providecommand \citenamefont [1]{#1}%
\providecommand \href@noop [0]{\@secondoftwo}%
\providecommand \href [0]{\begingroup \@sanitize@url \@href}%
\providecommand \@href[1]{\@@startlink{#1}\@@href}%
\providecommand \@@href[1]{\endgroup#1\@@endlink}%
\providecommand \@sanitize@url [0]{\catcode `\\12\catcode `\$12\catcode
  `\&12\catcode `\#12\catcode `\^12\catcode `\_12\catcode `\%12\relax}%
\providecommand \@@startlink[1]{}%
\providecommand \@@endlink[0]{}%
\providecommand \url  [0]{\begingroup\@sanitize@url \@url }%
\providecommand \@url [1]{\endgroup\@href {#1}{\urlprefix }}%
\providecommand \urlprefix  [0]{URL }%
\providecommand \Eprint [0]{\href }%
\providecommand \doibase [0]{http://dx.doi.org/}%
\providecommand \selectlanguage [0]{\@gobble}%
\providecommand \bibinfo  [0]{\@secondoftwo}%
\providecommand \bibfield  [0]{\@secondoftwo}%
\providecommand \translation [1]{[#1]}%
\providecommand \BibitemOpen [0]{}%
\providecommand \bibitemStop [0]{}%
\providecommand \bibitemNoStop [0]{.\EOS\space}%
\providecommand \EOS [0]{\spacefactor3000\relax}%
\providecommand \BibitemShut  [1]{\csname bibitem#1\endcsname}%
\let\auto@bib@innerbib\@empty
%</preamble>
\bibitem [{\citenamefont {Prigogine}(1977)}]{pri77nobel}%
  \BibitemOpen
  \bibfield  {author} {\bibinfo {author} {\bibfnamefont {I.}~\bibnamefont
  {Prigogine}},\ }\bibfield  {title} {\enquote {\bibinfo {title} {Time,
  structure, and fluctuations},}\ }\href@noop {} {\bibfield  {journal}
  {\bibinfo  {journal} {Nobel Lectures}\ } (\bibinfo {year}
  {1977})}\BibitemShut {NoStop}%
\bibitem [{\citenamefont {England}(2013)}]{eng13}%
  \BibitemOpen
  \bibfield  {author} {\bibinfo {author} {\bibfnamefont {J.~L.}\ \bibnamefont
  {England}},\ }\bibfield  {title} {\enquote {\bibinfo {title} {Statistical
  physics of self-replication},}\ }\href@noop {} {\bibfield  {journal}
  {\bibinfo  {journal} {J. Chem. Phys.}\ }\textbf {\bibinfo {volume} {139}},\
  \bibinfo {pages} {121923} (\bibinfo {year} {2013})}\BibitemShut {NoStop}%
\bibitem [{\citenamefont {England}(2015)}]{eng15}%
  \BibitemOpen
  \bibfield  {author} {\bibinfo {author} {\bibfnamefont {J.~L.}\ \bibnamefont
  {England}},\ }\bibfield  {title} {\enquote {\bibinfo {title} {Dissipative
  adaptation in driven self-assembly},}\ }\href@noop {} {\bibfield  {journal}
  {\bibinfo  {journal} {Nat. Nanotech.}\ }\textbf {\bibinfo {volume} {10}},\
  \bibinfo {pages} {919--923} (\bibinfo {year} {2015})}\BibitemShut {NoStop}%
\bibitem [{\citenamefont {Ruelle}(2017)}]{rue17}%
  \BibitemOpen
  \bibfield  {author} {\bibinfo {author} {\bibfnamefont {D.}~\bibnamefont
  {Ruelle}},\ }\bibfield  {title} {\enquote {\bibinfo {title} {The origin of
  life seen from the point of view of non-equilibrium statistical mechanics},}\
  }\href@noop {} {\bibfield  {journal} {\bibinfo  {journal} {arXiv:1701.08388}\
  } (\bibinfo {year} {2017})}\BibitemShut {NoStop}%
\bibitem [{\citenamefont {Lan}\ \emph {et~al.}(2012)\citenamefont {Lan},
  \citenamefont {Sartori}, \citenamefont {Neumann}, \citenamefont {Sourjik},\
  and\ \citenamefont {Tu}}]{lan12}%
  \BibitemOpen
  \bibfield  {author} {\bibinfo {author} {\bibfnamefont {G.}~\bibnamefont
  {Lan}}, \bibinfo {author} {\bibfnamefont {P.}~\bibnamefont {Sartori}},
  \bibinfo {author} {\bibfnamefont {S.}~\bibnamefont {Neumann}}, \bibinfo
  {author} {\bibfnamefont {V.}~\bibnamefont {Sourjik}}, \ and\ \bibinfo
  {author} {\bibfnamefont {Y.}~\bibnamefont {Tu}},\ }\bibfield  {title}
  {\enquote {\bibinfo {title} {The energy-speed-accuracy trade-off in sensory
  adaptation},}\ }\href@noop {} {\bibfield  {journal} {\bibinfo  {journal}
  {Nat. Phys.}\ }\textbf {\bibinfo {volume} {8}},\ \bibinfo {pages} {422--428}
  (\bibinfo {year} {2012})}\BibitemShut {NoStop}%
\bibitem [{\citenamefont {{ten Wolde}}(2012)}]{ten12}%
  \BibitemOpen
  \bibfield  {author} {\bibinfo {author} {\bibfnamefont {P.~R.}\ \bibnamefont
  {{ten Wolde}}},\ }\bibfield  {title} {\enquote {\bibinfo {title} {The price
  of accuracy},}\ }\href@noop {} {\bibfield  {journal} {\bibinfo  {journal}
  {Nat. Phys.}\ }\textbf {\bibinfo {volume} {8}},\ \bibinfo {pages} {361--362}
  (\bibinfo {year} {2012})}\BibitemShut {NoStop}%
\bibitem [{\citenamefont {Martyushev}\ and\ \citenamefont
  {Seleznev}(2006)}]{mar06}%
  \BibitemOpen
  \bibfield  {author} {\bibinfo {author} {\bibfnamefont {L.~M.}\ \bibnamefont
  {Martyushev}}\ and\ \bibinfo {author} {\bibfnamefont {V.~D.}\ \bibnamefont
  {Seleznev}},\ }\bibfield  {title} {\enquote {\bibinfo {title} {Maximum
  entropy production principle in physics, chemistry and biology},}\
  }\href@noop {} {\bibfield  {journal} {\bibinfo  {journal} {Phys. Rep.}\
  }\textbf {\bibinfo {volume} {426}},\ \bibinfo {pages} {1--45} (\bibinfo
  {year} {2006})}\BibitemShut {NoStop}%
\bibitem [{\citenamefont {Dunlap}(1999)}]{dun99}%
  \BibitemOpen
  \bibfield  {author} {\bibinfo {author} {\bibfnamefont {J.~C.}\ \bibnamefont
  {Dunlap}},\ }\bibfield  {title} {\enquote {\bibinfo {title} {Molecular bases
  for circadian clocks},}\ }\href@noop {} {\bibfield  {journal} {\bibinfo
  {journal} {Cell}\ }\textbf {\bibinfo {volume} {96}},\ \bibinfo {pages} {271
  -- 290} (\bibinfo {year} {1999})}\BibitemShut {NoStop}%
\bibitem [{\citenamefont {Gonze}\ \emph {et~al.}(2002)\citenamefont {Gonze},
  \citenamefont {Halloy},\ and\ \citenamefont {Goldbeter}}]{gon02}%
  \BibitemOpen
  \bibfield  {author} {\bibinfo {author} {\bibfnamefont {D.}~\bibnamefont
  {Gonze}}, \bibinfo {author} {\bibfnamefont {J.}~\bibnamefont {Halloy}}, \
  and\ \bibinfo {author} {\bibfnamefont {A.}~\bibnamefont {Goldbeter}},\
  }\bibfield  {title} {\enquote {\bibinfo {title} {Robustness of circadian
  rhythms with respect to molecular noise},}\ }\href@noop {} {\bibfield
  {journal} {\bibinfo  {journal} {Proc. Natl. Acad. Sci.}\ }\textbf {\bibinfo
  {volume} {99}},\ \bibinfo {pages} {673--678} (\bibinfo {year}
  {2002})}\BibitemShut {NoStop}%
\bibitem [{\citenamefont {Takahashi}(2016)}]{tak16}%
  \BibitemOpen
  \bibfield  {author} {\bibinfo {author} {\bibfnamefont {J.~S.}\ \bibnamefont
  {Takahashi}},\ }\bibfield  {title} {\enquote {\bibinfo {title}
  {Transcriptional architecture of the mammalian circadian clock},}\
  }\href@noop {} {\bibfield  {journal} {\bibinfo  {journal} {Nat. Rev. Genet.}\
  }\textbf {\bibinfo {volume} {18}},\ \bibinfo {pages} {164--179} (\bibinfo
  {year} {2016})}\BibitemShut {NoStop}%
\bibitem [{\citenamefont {Bi}\ \emph {et~al.}(2015)\citenamefont {Bi},
  \citenamefont {Lopez}, \citenamefont {Schwarz},\ and\ \citenamefont
  {Manning}}]{bi15}%
  \BibitemOpen
  \bibfield  {author} {\bibinfo {author} {\bibfnamefont {Dapeng}\ \bibnamefont
  {Bi}}, \bibinfo {author} {\bibfnamefont {J.~H.}\ \bibnamefont {Lopez}},
  \bibinfo {author} {\bibfnamefont {J.~M.}\ \bibnamefont {Schwarz}}, \ and\
  \bibinfo {author} {\bibfnamefont {M.~L.}\ \bibnamefont {Manning}},\
  }\bibfield  {title} {\enquote {\bibinfo {title} {A density-independent
  rigidity transition in biological tissues},}\ }\href@noop {} {\bibfield
  {journal} {\bibinfo  {journal} {Nat. Phys.}\ }\textbf {\bibinfo {volume}
  {11}},\ \bibinfo {pages} {1074--1079} (\bibinfo {year} {2015})}\BibitemShut
  {NoStop}%
\bibitem [{\citenamefont {Bi}\ \emph {et~al.}(2016)\citenamefont {Bi},
  \citenamefont {Yang}, \citenamefont {Marchetti},\ and\ \citenamefont
  {Manning}}]{bi16}%
  \BibitemOpen
  \bibfield  {author} {\bibinfo {author} {\bibfnamefont {Dapeng}\ \bibnamefont
  {Bi}}, \bibinfo {author} {\bibfnamefont {X.}~\bibnamefont {Yang}}, \bibinfo
  {author} {\bibfnamefont {M.~C.}\ \bibnamefont {Marchetti}}, \ and\ \bibinfo
  {author} {\bibfnamefont {M.~L.}\ \bibnamefont {Manning}},\ }\bibfield
  {title} {\enquote {\bibinfo {title} {Motility-driven glass and jamming
  transitions in biological tissues},}\ }\href@noop {} {\bibfield  {journal}
  {\bibinfo  {journal} {Phys. Rev. X}\ }\textbf {\bibinfo {volume} {6}},\
  \bibinfo {pages} {021011} (\bibinfo {year} {2016})}\BibitemShut {NoStop}%
\bibitem [{\citenamefont {Hoffmann}(2016)}]{hof16}%
  \BibitemOpen
  \bibfield  {author} {\bibinfo {author} {\bibfnamefont {P.~M.}\ \bibnamefont
  {Hoffmann}},\ }\bibfield  {title} {\enquote {\bibinfo {title} {How molecular
  motors extract order from chaos (a key issues review)},}\ }\href@noop {}
  {\bibfield  {journal} {\bibinfo  {journal} {Rep. Prog. Phys.}\ }\textbf
  {\bibinfo {volume} {79}},\ \bibinfo {pages} {032601} (\bibinfo {year}
  {2016})}\BibitemShut {NoStop}%
\bibitem [{\citenamefont {Lau}\ \emph {et~al.}(2007)\citenamefont {Lau},
  \citenamefont {Lacoste},\ and\ \citenamefont {Mallick}}]{lau07}%
  \BibitemOpen
  \bibfield  {author} {\bibinfo {author} {\bibfnamefont {A.~W.~C.}\
  \bibnamefont {Lau}}, \bibinfo {author} {\bibfnamefont {D.}~\bibnamefont
  {Lacoste}}, \ and\ \bibinfo {author} {\bibfnamefont {K.}~\bibnamefont
  {Mallick}},\ }\bibfield  {title} {\enquote {\bibinfo {title} {Nonequilibrium
  fluctuations and mechanochemical couplings of a molecular motor},}\
  }\href@noop {} {\bibfield  {journal} {\bibinfo  {journal} {Phys. Rev. Lett.}\
  }\textbf {\bibinfo {volume} {99}},\ \bibinfo {pages} {158102} (\bibinfo
  {year} {2007})}\BibitemShut {NoStop}%
\bibitem [{\citenamefont {Maes}\ and\ \citenamefont {{O'Kelly de
  Galway}}(2015)}]{mae15}%
  \BibitemOpen
  \bibfield  {author} {\bibinfo {author} {\bibfnamefont {C.}~\bibnamefont
  {Maes}}\ and\ \bibinfo {author} {\bibfnamefont {W.}~\bibnamefont {{O'Kelly de
  Galway}}},\ }\bibfield  {title} {\enquote {\bibinfo {title} {On the kinetics
  that moves {M}yosin {V}},}\ }\href@noop {} {\bibfield  {journal} {\bibinfo
  {journal} {Physica A}\ }\textbf {\bibinfo {volume} {436}} (\bibinfo {year}
  {2015})}\BibitemShut {NoStop}%
\bibitem [{\citenamefont {Parmeggiani}\ \emph {et~al.}(1999)\citenamefont
  {Parmeggiani}, \citenamefont {J\"ulicher}, \citenamefont {Ajdari},\ and\
  \citenamefont {Prost}}]{par99}%
  \BibitemOpen
  \bibfield  {author} {\bibinfo {author} {\bibfnamefont {A.}~\bibnamefont
  {Parmeggiani}}, \bibinfo {author} {\bibfnamefont {F.}~\bibnamefont
  {J\"ulicher}}, \bibinfo {author} {\bibfnamefont {A.}~\bibnamefont {Ajdari}},
  \ and\ \bibinfo {author} {\bibfnamefont {J.}~\bibnamefont {Prost}},\
  }\bibfield  {title} {\enquote {\bibinfo {title} {Energy transduction of
  isothermal ratchets: Generic aspects and specific examples close to and far
  from equilibrium},}\ }\href@noop {} {\bibfield  {journal} {\bibinfo
  {journal} {Phys. Rev. E}\ }\textbf {\bibinfo {volume} {60}},\ \bibinfo
  {pages} {2127--2140} (\bibinfo {year} {1999})}\BibitemShut {NoStop}%
\bibitem [{\citenamefont {Hopfield}(1974)}]{hop74}%
  \BibitemOpen
  \bibfield  {author} {\bibinfo {author} {\bibfnamefont {J.~J.}\ \bibnamefont
  {Hopfield}},\ }\bibfield  {title} {\enquote {\bibinfo {title} {Kinetic
  proofreading: A new mechanism for reducing errors in biosynthetic processes
  requiring high specificity},}\ }\href@noop {} {\bibfield  {journal} {\bibinfo
   {journal} {Proc. Natl. Acad. Sci.}\ }\textbf {\bibinfo {volume} {71}},\
  \bibinfo {pages} {4135--4139} (\bibinfo {year} {1974})}\BibitemShut {NoStop}%
\bibitem [{\citenamefont {Sartori}\ and\ \citenamefont
  {Pigolotti}(2013)}]{sar13}%
  \BibitemOpen
  \bibfield  {author} {\bibinfo {author} {\bibfnamefont {P.}~\bibnamefont
  {Sartori}}\ and\ \bibinfo {author} {\bibfnamefont {S.}~\bibnamefont
  {Pigolotti}},\ }\bibfield  {title} {\enquote {\bibinfo {title} {Kinetic
  versus energetic discrimination in biological copying},}\ }\href@noop {}
  {\bibfield  {journal} {\bibinfo  {journal} {Phys. Rev. Lett.}\ }\textbf
  {\bibinfo {volume} {110}},\ \bibinfo {pages} {188101} (\bibinfo {year}
  {2013})}\BibitemShut {NoStop}%
\bibitem [{\citenamefont {Gaspard}(2016)}]{gas16}%
  \BibitemOpen
  \bibfield  {author} {\bibinfo {author} {\bibfnamefont {P.}~\bibnamefont
  {Gaspard}},\ }\bibfield  {title} {\enquote {\bibinfo {title} {Kinetics and
  thermodynamics of {DNA} polymerases with exonuclease proofreading},}\
  }\href@noop {} {\bibfield  {journal} {\bibinfo  {journal} {Phys. Rev. E}\
  }\textbf {\bibinfo {volume} {93}},\ \bibinfo {pages} {042420} (\bibinfo
  {year} {2016})}\BibitemShut {NoStop}%
\bibitem [{\citenamefont {Ouldridge}\ \emph {et~al.}(2017)\citenamefont
  {Ouldridge}, \citenamefont {Govern},\ and\ \citenamefont {{ten
  Wolde}}}]{oul17}%
  \BibitemOpen
  \bibfield  {author} {\bibinfo {author} {\bibfnamefont {T.~E.}\ \bibnamefont
  {Ouldridge}}, \bibinfo {author} {\bibfnamefont {C.~C.}\ \bibnamefont
  {Govern}}, \ and\ \bibinfo {author} {\bibfnamefont {P.~R.}\ \bibnamefont
  {{ten Wolde}}},\ }\bibfield  {title} {\enquote {\bibinfo {title}
  {Thermodynamics of computational copying in biochemical systems},}\
  }\href@noop {} {\bibfield  {journal} {\bibinfo  {journal} {Phys. Rev. X}\
  }\textbf {\bibinfo {volume} {7}},\ \bibinfo {pages} {021004} (\bibinfo {year}
  {2017})}\BibitemShut {NoStop}%
\bibitem [{\citenamefont {Govern}\ and\ \citenamefont {{ten
  Wolde}}(2014)}]{gov14}%
  \BibitemOpen
  \bibfield  {author} {\bibinfo {author} {\bibfnamefont {C.~C.}\ \bibnamefont
  {Govern}}\ and\ \bibinfo {author} {\bibfnamefont {P.~R.}\ \bibnamefont {{ten
  Wolde}}},\ }\bibfield  {title} {\enquote {\bibinfo {title} {Optimal resource
  allocation in cellular sensing systems},}\ }\href@noop {} {\bibfield
  {journal} {\bibinfo  {journal} {Proc. Natl. Acad. Sci.}\ }\textbf {\bibinfo
  {volume} {111}},\ \bibinfo {pages} {17486--17491} (\bibinfo {year}
  {2014})}\BibitemShut {NoStop}%
\bibitem [{\citenamefont {Cui}\ and\ \citenamefont {Mehta}(2017)}]{cui17arxiv}%
  \BibitemOpen
  \bibfield  {author} {\bibinfo {author} {\bibfnamefont {W.}~\bibnamefont
  {Cui}}\ and\ \bibinfo {author} {\bibfnamefont {P.}~\bibnamefont {Mehta}},\
  }\bibfield  {title} {\enquote {\bibinfo {title} {Optimally in kinetic
  proofreading and early t-cell recognition: revisiting the speed, energy,
  accuracy trade-off},}\ }\href@noop {} {\bibfield  {journal} {\bibinfo
  {journal} {arXiv:1703.03398}\ } (\bibinfo {year} {2017})}\BibitemShut
  {NoStop}%
\bibitem [{\citenamefont {Deshpande}\ and\ \citenamefont
  {Ouldridge}(2017)}]{des17arxiv}%
  \BibitemOpen
  \bibfield  {author} {\bibinfo {author} {\bibfnamefont {A.}~\bibnamefont
  {Deshpande}}\ and\ \bibinfo {author} {\bibfnamefont {T.~E.}\ \bibnamefont
  {Ouldridge}},\ }\bibfield  {title} {\enquote {\bibinfo {title} {High rates of
  fuel consumption are not required by insulating motifs to suppress
  retroactivity in biochemical circuits},}\ }\href@noop {} {\bibfield
  {journal} {\bibinfo  {journal} {arXiv:1708.01792v3}\ } (\bibinfo {year}
  {2017})}\BibitemShut {NoStop}%
\bibitem [{\citenamefont {Hartl}(2016)}]{har16}%
  \BibitemOpen
  \bibfield  {author} {\bibinfo {author} {\bibfnamefont {F.~U.}\ \bibnamefont
  {Hartl}},\ }\bibfield  {title} {\enquote {\bibinfo {title} {Cellular
  homeostasis and aging},}\ }\href@noop {} {\bibfield  {journal} {\bibinfo
  {journal} {Annu. Rev. Biochem.}\ }\textbf {\bibinfo {volume} {85}},\ \bibinfo
  {pages} {1--4} (\bibinfo {year} {2016})}\BibitemShut {NoStop}%
\bibitem [{\citenamefont {{De Palo}}\ and\ \citenamefont
  {Endres}(2013)}]{dep13}%
  \BibitemOpen
  \bibfield  {author} {\bibinfo {author} {\bibfnamefont {G.}~\bibnamefont {{De
  Palo}}}\ and\ \bibinfo {author} {\bibfnamefont {R.~G.}\ \bibnamefont
  {Endres}},\ }\bibfield  {title} {\enquote {\bibinfo {title} {Unraveling
  adaptation in eukaryotic pathways: Lessons from protocells.}}\ }\href@noop {}
  {\bibfield  {journal} {\bibinfo  {journal} {PLoS Comp. Biol.}\ }\textbf
  {\bibinfo {volume} {9}} (\bibinfo {year} {2013})}\BibitemShut {NoStop}%
\bibitem [{\citenamefont {Buijsman}\ and\ \citenamefont
  {Sheinman}(2014)}]{bui14}%
  \BibitemOpen
  \bibfield  {author} {\bibinfo {author} {\bibfnamefont {W.}~\bibnamefont
  {Buijsman}}\ and\ \bibinfo {author} {\bibfnamefont {M.}~\bibnamefont
  {Sheinman}},\ }\bibfield  {title} {\enquote {\bibinfo {title} {Efficient
  fold-change detection based on protein-protein interactions},}\ }\href@noop
  {} {\bibfield  {journal} {\bibinfo  {journal} {Phys. Rev. E}\ }\textbf
  {\bibinfo {volume} {89}},\ \bibinfo {pages} {022712} (\bibinfo {year}
  {2014})}\BibitemShut {NoStop}%
\bibitem [{\citenamefont {Bruers}\ \emph {et~al.}(2007)\citenamefont {Bruers},
  \citenamefont {Maes},\ and\ \citenamefont {Neto\v{c}n\'{y}}}]{bru07}%
  \BibitemOpen
  \bibfield  {author} {\bibinfo {author} {\bibfnamefont {S.}~\bibnamefont
  {Bruers}}, \bibinfo {author} {\bibfnamefont {C.}~\bibnamefont {Maes}}, \ and\
  \bibinfo {author} {\bibfnamefont {K.}~\bibnamefont {Neto\v{c}n\'{y}}},\
  }\bibfield  {title} {\enquote {\bibinfo {title} {On the validity of entropy
  production principles for linear electrical circuits},}\ }\href@noop {}
  {\bibfield  {journal} {\bibinfo  {journal} {J. Stat. Phys.}\ }\textbf
  {\bibinfo {volume} {129}},\ \bibinfo {pages} {725--740} (\bibinfo {year}
  {2007})}\BibitemShut {NoStop}%
\bibitem [{\citenamefont {Landauer}(1975)}]{lan75}%
  \BibitemOpen
  \bibfield  {author} {\bibinfo {author} {\bibfnamefont {R.}~\bibnamefont
  {Landauer}},\ }\bibfield  {title} {\enquote {\bibinfo {title} {Inadequacy of
  entropy and entropy derivatives in characterizing the steady state},}\
  }\href@noop {} {\bibfield  {journal} {\bibinfo  {journal} {Phys. Rev. A}\
  }\textbf {\bibinfo {volume} {12}},\ \bibinfo {pages} {636--638} (\bibinfo
  {year} {1975})}\BibitemShut {NoStop}%
\bibitem [{\citenamefont {Maes}\ and\ \citenamefont
  {{Neto\v{c}n\'{y}}}(2013)}]{mae13.b}%
  \BibitemOpen
  \bibfield  {author} {\bibinfo {author} {\bibfnamefont {C}~\bibnamefont
  {Maes}}\ and\ \bibinfo {author} {\bibfnamefont {K}~\bibnamefont
  {{Neto\v{c}n\'{y}}}},\ }\bibfield  {title} {\enquote {\bibinfo {title} {Heat
  bounds and the blowtorch theorem},}\ }\href@noop {} {\bibfield  {journal}
  {\bibinfo  {journal} {Ann. Henri Poincar\'e}\ }\textbf {\bibinfo {volume}
  {14}},\ \bibinfo {pages} {1193--1202} (\bibinfo {year} {2013})}\BibitemShut
  {NoStop}%
\bibitem [{\citenamefont {Lecomte}\ \emph {et~al.}(2005)\citenamefont
  {Lecomte}, \citenamefont {Appert-Rolland},\ and\ \citenamefont {{van
  Wijland}}}]{lec05}%
  \BibitemOpen
  \bibfield  {author} {\bibinfo {author} {\bibfnamefont {V.}~\bibnamefont
  {Lecomte}}, \bibinfo {author} {\bibfnamefont {C.}~\bibnamefont
  {Appert-Rolland}}, \ and\ \bibinfo {author} {\bibfnamefont {F.}~\bibnamefont
  {{van Wijland}}},\ }\bibfield  {title} {\enquote {\bibinfo {title} {Chaotic
  properties of systems with {M}arkov dynamics},}\ }\href@noop {} {\bibfield
  {journal} {\bibinfo  {journal} {Phys. Rev. Lett.}\ }\textbf {\bibinfo
  {volume} {95}},\ \bibinfo {pages} {010601} (\bibinfo {year}
  {2005})}\BibitemShut {NoStop}%
\bibitem [{\citenamefont {Merolle}\ \emph {et~al.}(2005)\citenamefont
  {Merolle}, \citenamefont {Garrahan},\ and\ \citenamefont {Chandler}}]{mer05}%
  \BibitemOpen
  \bibfield  {author} {\bibinfo {author} {\bibfnamefont {M.}~\bibnamefont
  {Merolle}}, \bibinfo {author} {\bibfnamefont {J.~P.}\ \bibnamefont
  {Garrahan}}, \ and\ \bibinfo {author} {\bibfnamefont {D.}~\bibnamefont
  {Chandler}},\ }\bibfield  {title} {\enquote {\bibinfo {title} {Space-time
  thermodynampics of the glass transition},}\ }\href@noop {} {\bibfield
  {journal} {\bibinfo  {journal} {Proc. Natl. Acad. Sci.}\ }\textbf {\bibinfo
  {volume} {102}},\ \bibinfo {pages} {10837--10840} (\bibinfo {year}
  {2005})}\BibitemShut {NoStop}%
\bibitem [{\citenamefont {Garrahan}\ \emph {et~al.}(2009)\citenamefont
  {Garrahan}, \citenamefont {Jack}, \citenamefont {Lecomte}, \citenamefont
  {Pitard}, \citenamefont {{van Duijvendijk}},\ and\ \citenamefont {{van
  Wijland}}}]{gar09}%
  \BibitemOpen
  \bibfield  {author} {\bibinfo {author} {\bibfnamefont {J.~P.}\ \bibnamefont
  {Garrahan}}, \bibinfo {author} {\bibfnamefont {R.~L.}\ \bibnamefont {Jack}},
  \bibinfo {author} {\bibfnamefont {V.}~\bibnamefont {Lecomte}}, \bibinfo
  {author} {\bibfnamefont {E.}~\bibnamefont {Pitard}}, \bibinfo {author}
  {\bibfnamefont {K.}~\bibnamefont {{van Duijvendijk}}}, \ and\ \bibinfo
  {author} {\bibfnamefont {F.}~\bibnamefont {{van Wijland}}},\ }\bibfield
  {title} {\enquote {\bibinfo {title} {First-order dynamical phase transition
  in models of glasses: an approach based on ensembles of histories},}\
  }\href@noop {} {\bibfield  {journal} {\bibinfo  {journal} {J. Phys. A: Math.
  Gen}\ }\textbf {\bibinfo {volume} {42}},\ \bibinfo {pages} {075007} (\bibinfo
  {year} {2009})}\BibitemShut {NoStop}%
\bibitem [{\citenamefont {Gorissen}\ \emph {et~al.}(2009)\citenamefont
  {Gorissen}, \citenamefont {Hooyberghs},\ and\ \citenamefont
  {Vanderzande}}]{gor09}%
  \BibitemOpen
  \bibfield  {author} {\bibinfo {author} {\bibfnamefont {M.}~\bibnamefont
  {Gorissen}}, \bibinfo {author} {\bibfnamefont {J.}~\bibnamefont
  {Hooyberghs}}, \ and\ \bibinfo {author} {\bibfnamefont {C.}~\bibnamefont
  {Vanderzande}},\ }\bibfield  {title} {\enquote {\bibinfo {title}
  {Density-matrix renormalization-group study of current and activity
  fluctuations near nonequilibrium phase transitions},}\ }\href@noop {}
  {\bibfield  {journal} {\bibinfo  {journal} {Phys. Rev. E}\ }\textbf {\bibinfo
  {volume} {79}},\ \bibinfo {pages} {020101} (\bibinfo {year}
  {2009})}\BibitemShut {NoStop}%
\bibitem [{\citenamefont {Baiesi}\ \emph {et~al.}(2009)\citenamefont {Baiesi},
  \citenamefont {Maes},\ and\ \citenamefont {Wynants}}]{bai09}%
  \BibitemOpen
  \bibfield  {author} {\bibinfo {author} {\bibfnamefont {M.}~\bibnamefont
  {Baiesi}}, \bibinfo {author} {\bibfnamefont {C.}~\bibnamefont {Maes}}, \ and\
  \bibinfo {author} {\bibfnamefont {B.}~\bibnamefont {Wynants}},\ }\bibfield
  {title} {\enquote {\bibinfo {title} {Fluctuations and response of
  nonequilibrium states},}\ }\href@noop {} {\bibfield  {journal} {\bibinfo
  {journal} {Phys. Rev. Lett.}\ }\textbf {\bibinfo {volume} {103}},\ \bibinfo
  {pages} {010602} (\bibinfo {year} {2009})}\BibitemShut {NoStop}%
\bibitem [{\citenamefont {Baiesi}\ and\ \citenamefont {Maes}(2013)}]{bai13}%
  \BibitemOpen
  \bibfield  {author} {\bibinfo {author} {\bibfnamefont {M.}~\bibnamefont
  {Baiesi}}\ and\ \bibinfo {author} {\bibfnamefont {C.}~\bibnamefont {Maes}},\
  }\bibfield  {title} {\enquote {\bibinfo {title} {An update on the
  nonequilibrium linear response},}\ }\href@noop {} {\bibfield  {journal}
  {\bibinfo  {journal} {New J. Phys.}\ }\textbf {\bibinfo {volume} {15}},\
  \bibinfo {pages} {013004} (\bibinfo {year} {2013})}\BibitemShut {NoStop}%
\bibitem [{\citenamefont {Lippiello}\ \emph {et~al.}(2014)\citenamefont
  {Lippiello}, \citenamefont {Baiesi},\ and\ \citenamefont
  {Sarracino}}]{lip14}%
  \BibitemOpen
  \bibfield  {author} {\bibinfo {author} {\bibfnamefont {E.}~\bibnamefont
  {Lippiello}}, \bibinfo {author} {\bibfnamefont {M.}~\bibnamefont {Baiesi}}, \
  and\ \bibinfo {author} {\bibfnamefont {A.}~\bibnamefont {Sarracino}},\
  }\bibfield  {title} {\enquote {\bibinfo {title} {Nonequilibrium
  fluctuation-dissipation theorem and heat production},}\ }\href@noop {}
  {\bibfield  {journal} {\bibinfo  {journal} {Phys. Rev. Lett.}\ }\textbf
  {\bibinfo {volume} {112}},\ \bibinfo {pages} {140602} (\bibinfo {year}
  {2014})}\BibitemShut {NoStop}%
\bibitem [{\citenamefont {Basu}\ and\ \citenamefont {Maes}(2015)}]{bas15}%
  \BibitemOpen
  \bibfield  {author} {\bibinfo {author} {\bibfnamefont {U.}~\bibnamefont
  {Basu}}\ and\ \bibinfo {author} {\bibfnamefont {C.}~\bibnamefont {Maes}},\
  }\bibfield  {title} {\enquote {\bibinfo {title} {Nonequilibrium response and
  frenesy},}\ }\href@noop {} {\bibfield  {journal} {\bibinfo  {journal} {J. of
  Phys.: Conf. Ser.}\ }\textbf {\bibinfo {volume} {638}},\ \bibinfo {pages}
  {012001} (\bibinfo {year} {2015})}\BibitemShut {NoStop}%
\bibitem [{\citenamefont {Maes}(2016{\natexlab{a}})}]{mae16arxiv}%
  \BibitemOpen
  \bibfield  {author} {\bibinfo {author} {\bibfnamefont {C.}~\bibnamefont
  {Maes}},\ }\bibfield  {title} {\enquote {\bibinfo {title} {Nonequilibrium
  physics aspects of probabilistic cellular automata},}\ }\href@noop {}
  {\bibfield  {journal} {\bibinfo  {journal} {arXiv:1605.02876}\ } (\bibinfo
  {year} {2016}{\natexlab{a}})}\BibitemShut {NoStop}%
\bibitem [{\citenamefont {Maes}(2016{\natexlab{b}})}]{mae16}%
  \BibitemOpen
  \bibfield  {author} {\bibinfo {author} {\bibfnamefont {C.}~\bibnamefont
  {Maes}},\ }\bibfield  {title} {\enquote {\bibinfo {title} {What decides the
  direction of a current?}}\ }\href@noop {} {\bibfield  {journal} {\bibinfo
  {journal} {Math. Mech. Compl. Sys.}\ }\textbf {\bibinfo {volume} {3}},\
  \bibinfo {pages} {275--295} (\bibinfo {year}
  {2016}{\natexlab{b}})}\BibitemShut {NoStop}%
\bibitem [{\citenamefont {Jack}\ and\ \citenamefont {Evans}(2016)}]{jac16}%
  \BibitemOpen
  \bibfield  {author} {\bibinfo {author} {\bibfnamefont {R.~L.}\ \bibnamefont
  {Jack}}\ and\ \bibinfo {author} {\bibfnamefont {R.~M.~L.}\ \bibnamefont
  {Evans}},\ }\bibfield  {title} {\enquote {\bibinfo {title} {Absence of
  dissipation in trajectory ensembles biased by currents},}\ }\href@noop {}
  {\bibfield  {journal} {\bibinfo  {journal} {J. Stat. Mech.}\ ,\ \bibinfo
  {pages} {093305}} (\bibinfo {year} {2016})}\BibitemShut {NoStop}%
\bibitem [{\citenamefont {Helden}\ \emph {et~al.}(2016)\citenamefont {Helden},
  \citenamefont {Basu}, \citenamefont {Kr\"uger},\ and\ \citenamefont
  {Bechinger}}]{hel16}%
  \BibitemOpen
  \bibfield  {author} {\bibinfo {author} {\bibfnamefont {L.}~\bibnamefont
  {Helden}}, \bibinfo {author} {\bibfnamefont {U.}~\bibnamefont {Basu}},
  \bibinfo {author} {\bibfnamefont {M.}~\bibnamefont {Kr\"uger}}, \ and\
  \bibinfo {author} {\bibfnamefont {C.}~\bibnamefont {Bechinger}},\ }\bibfield
  {title} {\enquote {\bibinfo {title} {Measurement of second-order response
  without perturbation},}\ }\href@noop {} {\bibfield  {journal} {\bibinfo
  {journal} {Europhys. Lett.}\ }\textbf {\bibinfo {volume} {16}},\ \bibinfo
  {pages} {60003} (\bibinfo {year} {2016})}\BibitemShut {NoStop}%
\bibitem [{\citenamefont {Falasco}\ and\ \citenamefont
  {Baiesi}(2016)}]{fal16b}%
  \BibitemOpen
  \bibfield  {author} {\bibinfo {author} {\bibfnamefont {G.}~\bibnamefont
  {Falasco}}\ and\ \bibinfo {author} {\bibfnamefont {M.}~\bibnamefont
  {Baiesi}},\ }\bibfield  {title} {\enquote {\bibinfo {title} {Nonequilibrium
  temperature response for stochastic overdamped systems},}\ }\href@noop {}
  {\bibfield  {journal} {\bibinfo  {journal} {New J. Phys.}\ }\textbf {\bibinfo
  {volume} {18}},\ \bibinfo {pages} {043039} (\bibinfo {year}
  {2016})}\BibitemShut {NoStop}%
\bibitem [{\citenamefont {Yolcu}\ \emph {et~al.}(2017)\citenamefont {Yolcu},
  \citenamefont {B\'erut}, \citenamefont {Falasco}, \citenamefont {Petrosyan},
  \citenamefont {Ciliberto},\ and\ \citenamefont {Baiesi}}]{yol17}%
  \BibitemOpen
  \bibfield  {author} {\bibinfo {author} {\bibfnamefont {C.}~\bibnamefont
  {Yolcu}}, \bibinfo {author} {\bibfnamefont {A.}~\bibnamefont {B\'erut}},
  \bibinfo {author} {\bibfnamefont {G.}~\bibnamefont {Falasco}}, \bibinfo
  {author} {\bibfnamefont {A.}~\bibnamefont {Petrosyan}}, \bibinfo {author}
  {\bibfnamefont {S.}~\bibnamefont {Ciliberto}}, \ and\ \bibinfo {author}
  {\bibfnamefont {M.}~\bibnamefont {Baiesi}},\ }\bibfield  {title} {\enquote
  {\bibinfo {title} {A general fluctuation-response relation for noise
  variations and its application to driven hydrodynamic experiments},}\
  }\href@noop {} {\bibfield  {journal} {\bibinfo  {journal} {J. Stat. Phys.}\
  }\textbf {\bibinfo {volume} {167}},\ \bibinfo {pages} {29--45} (\bibinfo
  {year} {2017})}\BibitemShut {NoStop}%
\bibitem [{\citenamefont {Wark}\ \emph {et~al.}(2007)\citenamefont {Wark},
  \citenamefont {Lundstrom},\ and\ \citenamefont {Fairhall}}]{war07}%
  \BibitemOpen
  \bibfield  {author} {\bibinfo {author} {\bibfnamefont {B.}~\bibnamefont
  {Wark}}, \bibinfo {author} {\bibfnamefont {B.~N.}\ \bibnamefont {Lundstrom}},
  \ and\ \bibinfo {author} {\bibfnamefont {A.}~\bibnamefont {Fairhall}},\
  }\bibfield  {title} {\enquote {\bibinfo {title} {Sensory adaptation},}\
  }\href@noop {} {\bibfield  {journal} {\bibinfo  {journal} {Curr Opin
  Neurobiol.}\ }\textbf {\bibinfo {volume} {17}},\ \bibinfo {pages} {423--429}
  (\bibinfo {year} {2007})}\BibitemShut {NoStop}%
\bibitem [{\citenamefont {Derrida}(2007)}]{der07}%
  \BibitemOpen
  \bibfield  {author} {\bibinfo {author} {\bibfnamefont {B.}~\bibnamefont
  {Derrida}},\ }\bibfield  {title} {\enquote {\bibinfo {title} {Non-equilibrium
  steady states: fluctuations and large deviations of the density and of the
  current},}\ }\href@noop {} {\bibfield  {journal} {\bibinfo  {journal} {J.
  Stat. Mech.}\ ,\ \bibinfo {pages} {P07023}} (\bibinfo {year}
  {2007})}\BibitemShut {NoStop}%
\bibitem [{\citenamefont {Katz}\ \emph {et~al.}(1984)\citenamefont {Katz},
  \citenamefont {Lebowitz},\ and\ \citenamefont {Spohn}}]{kat84}%
  \BibitemOpen
  \bibfield  {author} {\bibinfo {author} {\bibfnamefont {S.}~\bibnamefont
  {Katz}}, \bibinfo {author} {\bibfnamefont {J.~L.}\ \bibnamefont {Lebowitz}},
  \ and\ \bibinfo {author} {\bibfnamefont {H.}~\bibnamefont {Spohn}},\
  }\bibfield  {title} {\enquote {\bibinfo {title} {Stationary nonequilibrium
  states for stochastic lattice gas models of ionic superconductors},}\
  }\href@noop {} {\bibfield  {journal} {\bibinfo  {journal} {J. Stat. Phys.}\
  }\textbf {\bibinfo {volume} {34}},\ \bibinfo {pages} {497} (\bibinfo {year}
  {1984})}\BibitemShut {NoStop}%
\bibitem [{\citenamefont {Maes}\ and\ \citenamefont
  {Neto\v{c}n\'{y}}(2003)}]{mae03}%
  \BibitemOpen
  \bibfield  {author} {\bibinfo {author} {\bibfnamefont {C.}~\bibnamefont
  {Maes}}\ and\ \bibinfo {author} {\bibfnamefont {K.}~\bibnamefont
  {Neto\v{c}n\'{y}}},\ }\bibfield  {title} {\enquote {\bibinfo {title}
  {Time-reversal and entropy},}\ }\href@noop {} {\bibfield  {journal} {\bibinfo
   {journal} {J. Stat. Phys.}\ }\textbf {\bibinfo {volume} {110}},\ \bibinfo
  {pages} {269--310} (\bibinfo {year} {2003})}\BibitemShut {NoStop}%
\bibitem [{\citenamefont {Harada}\ and\ \citenamefont {Sasa}(2005)}]{har05}%
  \BibitemOpen
  \bibfield  {author} {\bibinfo {author} {\bibfnamefont {T.}~\bibnamefont
  {Harada}}\ and\ \bibinfo {author} {\bibfnamefont {{S.-i}.}\ \bibnamefont
  {Sasa}},\ }\bibfield  {title} {\enquote {\bibinfo {title} {Equality
  connecting energy dissipation with violation of fluctuation-response
  relation},}\ }\href@noop {} {\bibfield  {journal} {\bibinfo  {journal} {Phys.
  Rev. Lett.}\ }\textbf {\bibinfo {volume} {95}},\ \bibinfo {pages} {130602}
  (\bibinfo {year} {2005})}\BibitemShut {NoStop}%
\bibitem [{\citenamefont {Crooks}(1998)}]{cro98}%
  \BibitemOpen
  \bibfield  {author} {\bibinfo {author} {\bibfnamefont {G.~E.}\ \bibnamefont
  {Crooks}},\ }\bibfield  {title} {\enquote {\bibinfo {title} {Nonequilibrium
  measurements of free energy differences for microscopically reversible
  {M}arkovian systems},}\ }\href@noop {} {\bibfield  {journal} {\bibinfo
  {journal} {J. Stat. Phys.}\ }\textbf {\bibinfo {volume} {90}},\ \bibinfo
  {pages} {1481} (\bibinfo {year} {1998})}\BibitemShut {NoStop}%
\bibitem [{\citenamefont {Maes}(1999)}]{mae99}%
  \BibitemOpen
  \bibfield  {author} {\bibinfo {author} {\bibfnamefont {C.}~\bibnamefont
  {Maes}},\ }\bibfield  {title} {\enquote {\bibinfo {title} {The fluctuation
  theorem as a {G}ibbs property},}\ }\href@noop {} {\bibfield  {journal}
  {\bibinfo  {journal} {J. Stat. Phys.}\ }\textbf {\bibinfo {volume} {95}},\
  \bibinfo {pages} {367--392} (\bibinfo {year} {1999})}\BibitemShut {NoStop}%
\bibitem [{\citenamefont {Maes}\ \emph {et~al.}(2000)\citenamefont {Maes},
  \citenamefont {Redig},\ and\ \citenamefont {{Van Moffaert}}}]{mae00}%
  \BibitemOpen
  \bibfield  {author} {\bibinfo {author} {\bibfnamefont {C.}~\bibnamefont
  {Maes}}, \bibinfo {author} {\bibfnamefont {F.}~\bibnamefont {Redig}}, \ and\
  \bibinfo {author} {\bibfnamefont {A.}~\bibnamefont {{Van Moffaert}}},\
  }\bibfield  {title} {\enquote {\bibinfo {title} {On the definition of entropy
  production, via examples},}\ }\href@noop {} {\bibfield  {journal} {\bibinfo
  {journal} {J. Mat. Phys.}\ }\textbf {\bibinfo {volume} {41}},\ \bibinfo
  {pages} {1528--1554} (\bibinfo {year} {2000})}\BibitemShut {NoStop}%
\bibitem [{\citenamefont {Maes}(2003)}]{maes03}%
  \BibitemOpen
  \bibfield  {author} {\bibinfo {author} {\bibfnamefont {C.}~\bibnamefont
  {Maes}},\ }\bibfield  {title} {\enquote {\bibinfo {title} {On the origin and
  the use of fluctuation relations for the entropy},}\ }\href@noop {}
  {\bibfield  {journal} {\bibinfo  {journal} {S\'eminaire Poincar\'e}\ }\textbf
  {\bibinfo {volume} {2}},\ \bibinfo {pages} {29--62} (\bibinfo {year}
  {2003})}\BibitemShut {NoStop}%
\bibitem [{\citenamefont {Maes}(2017)}]{mae17arxiv}%
  \BibitemOpen
  \bibfield  {author} {\bibinfo {author} {\bibfnamefont {C.}~\bibnamefont
  {Maes}},\ }\bibfield  {title} {\enquote {\bibinfo {title} {Frenetic bounds on
  the entropy production},}\ }\href@noop {} {\bibfield  {journal} {\bibinfo
  {journal} {arXiv:1705.07412}\ } (\bibinfo {year} {2017})}\BibitemShut
  {NoStop}%
\bibitem [{\citenamefont {Boksenbojm}\ and\ \citenamefont
  {Wynants}(2009)}]{bok09}%
  \BibitemOpen
  \bibfield  {author} {\bibinfo {author} {\bibfnamefont {E.}~\bibnamefont
  {Boksenbojm}}\ and\ \bibinfo {author} {\bibfnamefont {B.}~\bibnamefont
  {Wynants}},\ }\bibfield  {title} {\enquote {\bibinfo {title} {The entropy and
  efficiency of a molecular motor model},}\ }\href@noop {} {\bibfield
  {journal} {\bibinfo  {journal} {J. Phys. A: Math. Gen}\ }\textbf {\bibinfo
  {volume} {42}},\ \bibinfo {pages} {445003} (\bibinfo {year}
  {2009})}\BibitemShut {NoStop}%
\bibitem [{\citenamefont {Carter}\ and\ \citenamefont {Cross}(2006)}]{car06}%
  \BibitemOpen
  \bibfield  {author} {\bibinfo {author} {\bibfnamefont {N.~J.}\ \bibnamefont
  {Carter}}\ and\ \bibinfo {author} {\bibfnamefont {R.~A.}\ \bibnamefont
  {Cross}},\ }\bibfield  {title} {\enquote {\bibinfo {title} {Kinesin's
  moonwalk},}\ }\href@noop {} {\bibfield  {journal} {\bibinfo  {journal} {Curr.
  Opin. Cell Biol.}\ }\textbf {\bibinfo {volume} {18}},\ \bibinfo {pages}
  {61--67} (\bibinfo {year} {2006})}\BibitemShut {NoStop}%
\bibitem [{\citenamefont {Prigogine}\ and\ \citenamefont
  {Lefever}(1968)}]{pri68}%
  \BibitemOpen
  \bibfield  {author} {\bibinfo {author} {\bibfnamefont {I.}~\bibnamefont
  {Prigogine}}\ and\ \bibinfo {author} {\bibfnamefont {R.}~\bibnamefont
  {Lefever}},\ }\bibfield  {title} {\enquote {\bibinfo {title} {Symmetry
  breaking instabilities in dissipative systems. {II}},}\ }\href@noop {}
  {\bibfield  {journal} {\bibinfo  {journal} {J. Chem. Phys.}\ }\textbf
  {\bibinfo {volume} {48}},\ \bibinfo {pages} {1695--1700} (\bibinfo {year}
  {1968})}\BibitemShut {NoStop}%
\bibitem [{\citenamefont {Tyson}\ \emph {et~al.}(2008)\citenamefont {Tyson},
  \citenamefont {Albert}, \citenamefont {Goldbeter}, \citenamefont {Ruoff},\
  and\ \citenamefont {Sible}}]{tys08}%
  \BibitemOpen
  \bibfield  {author} {\bibinfo {author} {\bibfnamefont {J.~J.}\ \bibnamefont
  {Tyson}}, \bibinfo {author} {\bibfnamefont {R.}~\bibnamefont {Albert}},
  \bibinfo {author} {\bibfnamefont {A.}~\bibnamefont {Goldbeter}}, \bibinfo
  {author} {\bibfnamefont {P.}~\bibnamefont {Ruoff}}, \ and\ \bibinfo {author}
  {\bibfnamefont {J.}~\bibnamefont {Sible}},\ }\bibfield  {title} {\enquote
  {\bibinfo {title} {Biological switches and clocks},}\ }\href@noop {}
  {\bibfield  {journal} {\bibinfo  {journal} {J. Royal Soc. Interf.}\ }\textbf
  {\bibinfo {volume} {5(Suppl 1)}},\ \bibinfo {pages} {S1--S8} (\bibinfo {year}
  {2008})}\BibitemShut {NoStop}%
\bibitem [{\citenamefont {Wang}\ \emph {et~al.}(2015)\citenamefont {Wang},
  \citenamefont {Lan},\ and\ \citenamefont {Tang}}]{wan15}%
  \BibitemOpen
  \bibfield  {author} {\bibinfo {author} {\bibfnamefont {Shou-Wen}\
  \bibnamefont {Wang}}, \bibinfo {author} {\bibfnamefont {Yueheng}\
  \bibnamefont {Lan}}, \ and\ \bibinfo {author} {\bibfnamefont {Lei-Han}\
  \bibnamefont {Tang}},\ }\bibfield  {title} {\enquote {\bibinfo {title}
  {Energy dissipation in an adaptive molecular circuit},}\ }\href@noop {}
  {\bibfield  {journal} {\bibinfo  {journal} {J. Stat. Mech.}\ ,\ \bibinfo
  {pages} {P07025}} (\bibinfo {year} {2015})}\BibitemShut {NoStop}%
\bibitem [{\citenamefont {Sekimoto}(2010)}]{sek10}%
  \BibitemOpen
  \bibfield  {author} {\bibinfo {author} {\bibfnamefont {K.}~\bibnamefont
  {Sekimoto}},\ }\href@noop {} {\emph {\bibinfo {title} {Stochastic
  Energetics}}},\ \bibinfo {series} {Lecture Notes in Physics}, Vol.\ \bibinfo
  {volume} {799}\ (\bibinfo  {publisher} {Springer},\ \bibinfo {year}
  {2010})\BibitemShut {NoStop}%
\bibitem [{\citenamefont {Maes}\ \emph {et~al.}(2008)\citenamefont {Maes},
  \citenamefont {{Neto\v{c}n\'{y}}},\ and\ \citenamefont {Wynants}}]{mae08a}%
  \BibitemOpen
  \bibfield  {author} {\bibinfo {author} {\bibfnamefont {C.}~\bibnamefont
  {Maes}}, \bibinfo {author} {\bibfnamefont {K.}~\bibnamefont
  {{Neto\v{c}n\'{y}}}}, \ and\ \bibinfo {author} {\bibfnamefont
  {B.}~\bibnamefont {Wynants}},\ }\bibfield  {title} {\enquote {\bibinfo
  {title} {Steady state statistics of driven diffusions},}\ }\href@noop {}
  {\bibfield  {journal} {\bibinfo  {journal} {Physica A}\ }\textbf {\bibinfo
  {volume} {387}},\ \bibinfo {pages} {2675--2689} (\bibinfo {year}
  {2008})}\BibitemShut {NoStop}%
\bibitem [{\citenamefont {Baerts}\ \emph {et~al.}(2013)\citenamefont {Baerts},
  \citenamefont {Basu}, \citenamefont {Maes},\ and\ \citenamefont
  {Safaverdi}}]{bae13}%
  \BibitemOpen
  \bibfield  {author} {\bibinfo {author} {\bibfnamefont {P.}~\bibnamefont
  {Baerts}}, \bibinfo {author} {\bibfnamefont {U.}~\bibnamefont {Basu}},
  \bibinfo {author} {\bibfnamefont {C.}~\bibnamefont {Maes}}, \ and\ \bibinfo
  {author} {\bibfnamefont {S.}~\bibnamefont {Safaverdi}},\ }\bibfield  {title}
  {\enquote {\bibinfo {title} {Frenetic origin of negative differential
  response},}\ }\href@noop {} {\bibfield  {journal} {\bibinfo  {journal} {Phys.
  Rev. E}\ }\textbf {\bibinfo {volume} {88}},\ \bibinfo {pages} {052109}
  (\bibinfo {year} {2013})}\BibitemShut {NoStop}%
\bibitem [{\citenamefont {Baiesi}\ \emph {et~al.}(2015)\citenamefont {Baiesi},
  \citenamefont {Stella},\ and\ \citenamefont {Vanderzande}}]{bai15}%
  \BibitemOpen
  \bibfield  {author} {\bibinfo {author} {\bibfnamefont {M.}~\bibnamefont
  {Baiesi}}, \bibinfo {author} {\bibfnamefont {A.~L.}\ \bibnamefont {Stella}},
  \ and\ \bibinfo {author} {\bibfnamefont {C.}~\bibnamefont {Vanderzande}},\
  }\bibfield  {title} {\enquote {\bibinfo {title} {Role of trapping and
  crowding as sources of negative differential mobility},}\ }\href@noop {}
  {\bibfield  {journal} {\bibinfo  {journal} {Phys. Rev. E}\ }\textbf {\bibinfo
  {volume} {92}},\ \bibinfo {pages} {042121} (\bibinfo {year}
  {2015})}\BibitemShut {NoStop}%
\bibitem [{\citenamefont {Sarracino}\ \emph {et~al.}((2016)\citenamefont
  {Sarracino}, \citenamefont {Cecconi}, \citenamefont {Puglisi},\ and\
  \citenamefont {Vulpiani}}]{sar16}%
  \BibitemOpen
  \bibfield  {author} {\bibinfo {author} {\bibfnamefont {A.}~\bibnamefont
  {Sarracino}}, \bibinfo {author} {\bibfnamefont {F.}~\bibnamefont {Cecconi}},
  \bibinfo {author} {\bibfnamefont {A.}~\bibnamefont {Puglisi}}, \ and\
  \bibinfo {author} {\bibfnamefont {A.}~\bibnamefont {Vulpiani}},\ }\bibfield
  {title} {\enquote {\bibinfo {title} {Nonlinear response of inertial tracers
  in steady laminar flows: Differential and absolute negative mobility},}\
  }\href@noop {} {\bibfield  {journal} {\bibinfo  {journal} {Phys. Rev. Lett.}\
  } (\bibinfo {year} {(2016})}\BibitemShut {NoStop}%
\bibitem [{\citenamefont {Altaner}\ \emph {et~al.}(2016)\citenamefont
  {Altaner}, \citenamefont {Polettini},\ and\ \citenamefont
  {Esposito}}]{alt16}%
  \BibitemOpen
  \bibfield  {author} {\bibinfo {author} {\bibfnamefont {B.}~\bibnamefont
  {Altaner}}, \bibinfo {author} {\bibfnamefont {M.}~\bibnamefont {Polettini}},
  \ and\ \bibinfo {author} {\bibfnamefont {M.}~\bibnamefont {Esposito}},\
  }\bibfield  {title} {\enquote {\bibinfo {title} {Fluctuation-dissipation
  relations far from equilibrium},}\ }\href@noop {} {\bibfield  {journal}
  {\bibinfo  {journal} {Phys. Rev. Lett.}\ }\textbf {\bibinfo {volume} {117}},\
  \bibinfo {pages} {180601} (\bibinfo {year} {2016})}\BibitemShut {NoStop}%
\bibitem [{\citenamefont {Basu}\ \emph {et~al.}(2015)\citenamefont {Basu},
  \citenamefont {Kr\"uger}, \citenamefont {Lazarescu},\ and\ \citenamefont
  {Maes}}]{bas15.c}%
  \BibitemOpen
  \bibfield  {author} {\bibinfo {author} {\bibfnamefont {U.}~\bibnamefont
  {Basu}}, \bibinfo {author} {\bibfnamefont {M.}~\bibnamefont {Kr\"uger}},
  \bibinfo {author} {\bibfnamefont {A.}~\bibnamefont {Lazarescu}}, \ and\
  \bibinfo {author} {\bibfnamefont {C.}~\bibnamefont {Maes}},\ }\bibfield
  {title} {\enquote {\bibinfo {title} {Frenetic aspects of second order
  response},}\ }\href@noop {} {\bibfield  {journal} {\bibinfo  {journal} {Phys.
  Chem. Chem. Phys.}\ }\textbf {\bibinfo {volume} {17}},\ \bibinfo {pages}
  {6653--6666} (\bibinfo {year} {2015})}\BibitemShut {NoStop}%
\bibitem [{\citenamefont {Basu}\ \emph {et~al.}(2016)\citenamefont {Basu},
  \citenamefont {{de Buyl}}, \citenamefont {Maes},\ and\ \citenamefont
  {Neto\v{c}n\'{y}}}]{bas16}%
  \BibitemOpen
  \bibfield  {author} {\bibinfo {author} {\bibfnamefont {U.}~\bibnamefont
  {Basu}}, \bibinfo {author} {\bibfnamefont {P.}~\bibnamefont {{de Buyl}}},
  \bibinfo {author} {\bibfnamefont {C.}~\bibnamefont {Maes}}, \ and\ \bibinfo
  {author} {\bibfnamefont {K.}~\bibnamefont {Neto\v{c}n\'{y}}},\ }\bibfield
  {title} {\enquote {\bibinfo {title} {Driving-induced stability with
  long-range effects},}\ }\href@noop {} {\bibfield  {journal} {\bibinfo
  {journal} {Europhys. Lett.}\ }\textbf {\bibinfo {volume} {115}},\ \bibinfo
  {pages} {30007} (\bibinfo {year} {2016})}\BibitemShut {NoStop}%
\bibitem [{\citenamefont {Barato}\ and\ \citenamefont {Seifert}(2015)}]{bar15}%
  \BibitemOpen
  \bibfield  {author} {\bibinfo {author} {\bibfnamefont {A.~C.}\ \bibnamefont
  {Barato}}\ and\ \bibinfo {author} {\bibfnamefont {U.}~\bibnamefont
  {Seifert}},\ }\bibfield  {title} {\enquote {\bibinfo {title} {Thermodynamic
  uncertainty relation for biomolecular processes},}\ }\href@noop {} {\bibfield
   {journal} {\bibinfo  {journal} {Phys. Rev. Lett.}\ }\textbf {\bibinfo
  {volume} {114}},\ \bibinfo {pages} {158101} (\bibinfo {year}
  {2015})}\BibitemShut {NoStop}%
\bibitem [{\citenamefont {Polettini}\ \emph {et~al.}(2016)\citenamefont
  {Polettini}, \citenamefont {Lazarescu},\ and\ \citenamefont
  {Esposito}}]{pol16}%
  \BibitemOpen
  \bibfield  {author} {\bibinfo {author} {\bibfnamefont {M.}~\bibnamefont
  {Polettini}}, \bibinfo {author} {\bibfnamefont {A.}~\bibnamefont
  {Lazarescu}}, \ and\ \bibinfo {author} {\bibfnamefont {M.}~\bibnamefont
  {Esposito}},\ }\bibfield  {title} {\enquote {\bibinfo {title} {Tightening the
  uncertainty principle for stochastic currents},}\ }\href@noop {} {\bibfield
  {journal} {\bibinfo  {journal} {Phys. Rev. E}\ }\textbf {\bibinfo {volume}
  {94}},\ \bibinfo {pages} {052104} (\bibinfo {year} {2016})}\BibitemShut
  {NoStop}%
\bibitem [{\citenamefont {Pietzonka}\ \emph {et~al.}(2017)\citenamefont
  {Pietzonka}, \citenamefont {Ritort},\ and\ \citenamefont {Seifert}}]{pie17}%
  \BibitemOpen
  \bibfield  {author} {\bibinfo {author} {\bibfnamefont {P.}~\bibnamefont
  {Pietzonka}}, \bibinfo {author} {\bibfnamefont {F.}~\bibnamefont {Ritort}}, \
  and\ \bibinfo {author} {\bibfnamefont {U.}~\bibnamefont {Seifert}},\
  }\bibfield  {title} {\enquote {\bibinfo {title} {Finite-time generalization
  of the thermodynamic uncertainty relation},}\ }\href@noop {} {\bibfield
  {journal} {\bibinfo  {journal} {Phys. Rev. E}\ }\textbf {\bibinfo {volume}
  {96}},\ \bibinfo {pages} {012101} (\bibinfo {year} {2017})}\BibitemShut
  {NoStop}%
\bibitem [{\citenamefont {Pigolotti}\ \emph {et~al.}(2017)\citenamefont
  {Pigolotti}, \citenamefont {Neri}, \citenamefont {Rold\'an},\ and\
  \citenamefont {J\"ulicher}}]{pig17arxiv}%
  \BibitemOpen
  \bibfield  {author} {\bibinfo {author} {\bibfnamefont {S.}~\bibnamefont
  {Pigolotti}}, \bibinfo {author} {\bibfnamefont {I.}~\bibnamefont {Neri}},
  \bibinfo {author} {\bibfnamefont {{\'E}}~\bibnamefont {Rold\'an}}, \ and\
  \bibinfo {author} {\bibfnamefont {F.}~\bibnamefont {J\"ulicher}},\ }\bibfield
   {title} {\enquote {\bibinfo {title} {Generic properties of stochastic
  entropy production},}\ }\href@noop {} {\bibfield  {journal} {\bibinfo
  {journal} {arXiv:1704.04061}\ } (\bibinfo {year} {2017})}\BibitemShut
  {NoStop}%
\bibitem [{\citenamefont {Horowitz}\ and\ \citenamefont
  {Gingrich}(2017)}]{hor17}%
  \BibitemOpen
  \bibfield  {author} {\bibinfo {author} {\bibfnamefont {J.~M.}\ \bibnamefont
  {Horowitz}}\ and\ \bibinfo {author} {\bibfnamefont {T.~R.}\ \bibnamefont
  {Gingrich}},\ }\bibfield  {title} {\enquote {\bibinfo {title} {Proof of the
  finite-time thermodynamic uncertainty relation for steady-state currents},}\
  }\href@noop {} {\bibfield  {journal} {\bibinfo  {journal} {Phys. Rev. E}\
  }\textbf {\bibinfo {volume} {96}},\ \bibinfo {pages} {020103(R)} (\bibinfo
  {year} {2017})}\BibitemShut {NoStop}%
\bibitem [{\citenamefont {Seifert}(2017)}]{sei17arxiv}%
  \BibitemOpen
  \bibfield  {author} {\bibinfo {author} {\bibfnamefont {U.}~\bibnamefont
  {Seifert}},\ }\bibfield  {title} {\enquote {\bibinfo {title} {Stochastic
  thermodynamics: From principles to the cost of precision},}\ }\href@noop {}
  {\bibfield  {journal} {\bibinfo  {journal} {arXiv:1707.03759}\ } (\bibinfo
  {year} {2017})}\BibitemShut {NoStop}%
\bibitem [{\citenamefont {Horowitz}\ and\ \citenamefont
  {England}(2017)}]{hor17arxiv}%
  \BibitemOpen
  \bibfield  {author} {\bibinfo {author} {\bibfnamefont {J.~M.}\ \bibnamefont
  {Horowitz}}\ and\ \bibinfo {author} {\bibfnamefont {J.~L.}\ \bibnamefont
  {England}},\ }\bibfield  {title} {\enquote {\bibinfo {title}
  {Information-theoretic bound on the entropy production to maintain a
  classical nonequilibrium distribution using ancillary control},}\ }\href@noop
  {} {\bibfield  {journal} {\bibinfo  {journal} {arXiv:1707.00367}\ } (\bibinfo
  {year} {2017})}\BibitemShut {NoStop}%
\end{thebibliography}


\end{document}